\documentclass[draft,aps,pre,twocolumn,groupaddress,showkeys,nofootinbib,preprintnumbers,floatfix]{revtex4-2}

\usepackage{dynlearn}
\usepackage{cancel}
\usepackage{probs}
\usepackage{dsfont}
\usepackage{braket}

\newcommand{\pst}[1]{\overleftarrow{#1}}
\newcommand{\fut}[1]{\overrightarrow{#1}}

\begin{document}

\def\ourTitle{Topology, Convergence, and Reconstruction
  of Predictive States
}

\def\ourAbstract{Predictive equivalence in discrete stochastic processes have been applied with
great success to identify randomness and structure in statistical physics and
chaotic dynamical systems and to inferring hidden Markov models. We examine the
conditions under which they can be reliably reconstructed from time-series data,
showing that convergence of predictive states can be achieved from empirical
samples in the weak topology of measures. Moreover, predictive states may be
represented in Hilbert spaces that replicate the weak topology. We
mathematically explain how these representations are particularly beneficial
when reconstructing high-memory processes and connect them to reproducing
kernel Hilbert spaces.
}

\def\ourKeywords{stochastic process, symbolic dynamics, dynamical systems, measure theory, weak topology
}

\hypersetup{
  pdfauthor={James P. Crutchfield},
  pdftitle={\ourTitle},
  pdfsubject={\ourAbstract},
  pdfkeywords={\ourKeywords},
  pdfproducer={},
  pdfcreator={}
}

\author{Samuel P. Loomis}
\email{sloomis@ucdavis.edu}

\author{James P. Crutchfield}
\email{chaos@ucdavis.edu}

\affiliation{Complexity Sciences Center and Department of Physics and Astronomy,
University of California at Davis, One Shields Avenue, Davis, CA 95616}

\date{\today}
\bibliographystyle{unsrt}

\title{\ourTitle}

\begin{abstract}
\ourAbstract
\end{abstract}


\preprint{\arxiv{2109.XXXXX}}

\date{\today}
\maketitle

\makeatletter
\let\toc@pre\relax
\let\toc@post\relax
\makeatother

\setstretch{1.1}

\listoffixmes

\section{Introduction}

With an accurate model in hand, an observer can leverage their knowledge of a
system's history to predict its future behavior. For stochastic
processes---distributions over time-series data---the task of predicting future
behavior from past observations and the associated resource constraints this
task imposes on an observer have been studied under the physics of
\emph{computational mechanics} \cite{Crut12a}. This subfield of
statistical mechanics focuses on the intrinsic information-processing embedded
in natural systems.

Its chief insight is the concept of the predictive (or causal) state. A
process' predictive states play a dual role. On the one hand, to accurately
predict a process' future behavior they are the key objects that an observer
must be capable of reproducing in their model. On the other hand, the
predictive states and their dynamics are central to understanding the
intrinsic, model-independent properties of the process itself \cite{Crut12a}.

The concept of predictive states has found use in numerous settings, such as
classical and quantum thermodynamics \cite{Boyd14b,Loom19a,Jurg20a},
quantum information and computing \cite{Gu12a,Maho15a,Bind17a,Vene19a},
condensed matter \cite{Crut97a,Varn14a,Marz18a}, dynamical systems
\cite{Jurg20c}, cellular automata \cite{Rupe17b}, and
model inference
\cite{Shal02a,Stil07b,Stre13a,Marz14f,Rupe19b,Rupe20a,Brod20a}.
Additionally, in the setting of processes generated by finite-state,
discrete-output hidden Markov models (HMMs) and generalized hidden Markov
models (GHMMs), a deep mathematical theory of predictive states is now
available \cite{Uppe97a,Crut12a,Jaeg00a,Jame04a,Trav10b,Thon15a}.

Despite their broad utility, a mathematically rigorous definition of predictive
states is needed that is applicable and useful for even more general stochastic
processes. Here, we have in mind large-memory processes whose long-time
correlations cannot be finitely represented by HMMs or GHMMs and processes
whose outputs may span a continuous domain in time and space.

The following takes the next major step towards a rigorous and mathematically
general definition of predictive states, extending the concept to all processes
whose observations are temporally discrete but may otherwise be either discrete
or continuous.

Somewhat remarkably, for any stationary and ergodic stochastic process of this
kind, predictive states are always well-defined and, furthermore, may always be
convergently approximated from empirical observations given a sufficiently large
sample. Next, we expand on recent work on Hilbert space embeddings of predictive
states \cite{Song09a,Song10a,Boot13a,Brod20a}, demonstrate that such
embeddings always exist, and discuss their implications for predictive-state
geometry and topology. Last, we explore the implications of our results for
empirically reconstructing predictive states via reproducing kernel Hilbert
spaces, particularly through the addition of new terms in the asymptotic
convergence bounds.

\section{Assumptions and preliminaries}
\label{sec:prelim}
\subsection{Stochastic processes}
\label{sec:stochastic}

We begin by laying out a series of definitions and identifying the assumptions
made. We draw from the combined literature of measures, stochastic processes,
and symbolic dynamics \cite{Kall01a,Kurk03a}.

A \emph{stochastic process} is typically defined as a function-valued random
variable $X:\Omega\rightarrow \mathcal{X}^\mathcal{T}$, where
$(\Omega,\Sigma,\mu)$ is a measure space, $\mathcal{T}$ is a set of temporal
indices (perhaps the real line, perhaps a discrete set), and $\mathcal{X}$ is a
set of possible observations (also potentially real or discrete in nature). We
take the sample space $\Omega$ to be the set $\mathcal{X}^\mathcal{T}$ and $X$
to be the identity. In this way, a stochastic process is identified solely with
the measure $\mu$ over $\Omega=\mathcal{X}^\mathcal{T}$.

When $\mathcal{T}$ is $\mathbb{Z}$, we say the process is \emph{discrete-time};
when it is $\mathbb{R}$ we say \emph{continuous-time}. Unless specified
otherwise we {assume} discrete-time, later treating continuous-time as an
extension of the discrete case. In discrete time, it is convenient to write
$X(t)$ as an indexed sequence $(x_t)$, where each $x_t$ is an element of
$\mathcal{X}$. When $\mathcal{X}$ is a discrete finite set, we say that the
process is \emph{discrete-observation}; by \emph{continuous-observation} we
typically mean the case where $\mathcal{X}$ is an interval in $\mathbb{R}$ or a
Cartesian product of intervals in $\mathbb{R}^d$. These are the only cases we
consider rigorously.  That said, we believe they are sufficient for many
practical purposes or, at least, not too cumbersome to extend if necessary.

The temporal \emph{shift operator} $\tau:\mathcal{X}^\mathcal{T}
\rightarrow\mathcal{X}^\mathcal{T}$ simply translates $t \mapsto t+1$: $(\tau
X)(t) = X(t+1)$. It also acts on measures of $\mathcal{X}^\mathcal{T}$: $(\tau
\mu)(A)=\mu(\tau^{-1} A)$. A stochastic process paired with the shift
operator---$(\mathcal{X}^\mathcal{T},\Sigma,\mu,\tau)$---becomes a dynamical
system and is \emph{stationary} if $\tau\mu=\mu$. It is further considered
\emph{ergodic} if, for all shift-invariant sets
$\mathcal{I}\subseteq\mathcal{X}^\mathcal{T}$, either $\mu(\mathcal{I})=1$ or
$\mu(\mathcal{I})=0$. Here, we assume all processes are both stationary and
ergodic.

If $\mathcal{X}$ is discrete, then the measurable sets of
$\mathcal{X}^\mathcal{Z}$ are generated by the \emph{cylinder sets}:
\begin{align*}
    U_{t,w} := \Set{
        X: x_{t+1}\dots x_{t+\ell} = w
    }
	~,
\end{align*}
where $w\in \mathcal{X}^\ell$ is a \emph{word} of length $\ell$. For a
stationary process, the \emph{word probabilities}:
\begin{align*}
    \Prob{\mu}{x_1 \dots x_\ell} := \mu\left(U_{0,x_1\dots x_\ell}\right)
\end{align*}
are sufficient to uniquely define the measure $\mu$.

In the continuous-observation case, the issue is more subtle. A cylinder set
instead takes the form:

\begin{align*}
    U_{t,I_1\dots I_\ell} := \Set{
        X: x_{t+1}\in I_1,\dots,x_{t+\ell}\in I_\ell
    }
	~,
\end{align*}
where each $I_t$ is an interval in $\mathcal{X}$. This does not lend itself
well to expressing simple word probabilities. However, we can define the
\emph{word measures} $\mu_\ell$ by restricting $\mu$ to the set
$\mathcal{X}^\ell$ describing the first $\ell$ values.

\section{Predictive states}
\label{sec:predictive}

Each element $X\in\mathcal{X}^\mathbb{Z}$ can be decomposed from a bidirectional infinite sequence
to a pair of unidirectional infinite sequences in $\mathcal{X}^\mathbb{N}\times
\mathcal{X}^\mathbb{N}$, by the transformation $\dots x_{-1}x_0 x_1\dots \mapsto
(x_{0}x_{-1}\dots,x_1 x_2 \dots)$. The first sequence in this pair we call the
\emph{past} $\pst{X}$ and the second we call the \emph{future} $\fut{X}$. In
this perspective, a stochastic process is a bipartite measure over pasts and
futures. The intuitive definition of a \emph{predictive state} is as a measure
over future sequences that arises from conditioning on past sequences.
Heuristically, $\Prob{\mu}{\fut{X}|\pst{X}= x_0 x_{-1} \dots}$ represents the
``predictive state'' associated with past $x_0 x_{-1}\dots$.

Conditioning of measures is a nuanced issue, especially when the involved
sample spaces are uncountably infinite \cite{Rao05a}. Of the many perspectives
that define a conditional measure, the most practical and intuitive is that a
conditional measure is a ratio of likelihoods---and, in the continuous case, a
limit of such ratios. However, determining the manner in which this limit must
be taken is rarely trivial. The following considers first the case of discrete
observations, where the matter is relatively straightforward. Then we examine
the case of continuous observations, reviewing the previous literature on the
nuances of this domain and extending its results for our present purposes. As
we will see, in either case, the intuition of predictive states can be born out
in a rigorous and elegant manner for any stochastic process satisfying the
assumptions heretofore mentioned.

\subsection{Discrete observations}
\label{sec:discrete}

We first establish likelihood-ratio convergence.

\begin{The}
\label{thm:discrete-pstate}
For all measures $\mu$ on $\mathcal{X}^\mathbb{Z}$, all $\ell\in\mathbb{N}$, all
$w = x_1 \dots x_\ell \in\mathcal{X}^\ell$, and $\pst{\mu}$-almost all pasts
$\pst{X}$, where $\mathcal{X}$ is a finite set, the limit:
\begin{align}
\label{eq:discrete-pstate-def}
    \Prob{\mu}{w|\pst{X}} := 
    \lim_{k\rightarrow \infty}
    \frac{\Prob{\mu}{x_{-k}\dots x_0 x_1 \dots x_\ell}}
    {\Prob{\mu}{x_{-k}\dots x_0}}
\end{align}
is convergent.
\end{The}

For all $\pst{X}$ where \cref{eq:discrete-pstate-def} converges, we can define a
measure $\epsilon[\pst{X}]\in\mathbb{M}(\mathcal{X}^\mathbb{N})$ over future
sequences, uniquely determined by the requirement $\epsilon[\pst{X}](U_{0,w}) =
\Prob{\mu}{w|\pst{X}}$. This $\epsilon[\pst{X}]$ is the \emph{predictive
state} of $\pst{X}$ and the function $\epsilon:\mathcal{X}^\mathbb{N}\rightarrow
\mathbb{M}(\mathcal{X}^\mathbb{N})$, the \emph{prediction mapping}.

The proof strategy consists in redefining the problem. The limit
\cref{eq:discrete-pstate-def} can be recast as what is called a
\emph{likelihood ratio}. The convergence of likelihood ratios is itself closely
related to the theory of Radon-Nikodym derivatives between measures.
Specifically, the Radon-Nikodym derivative can be computed as a convergence of
likelihood ratios. That convergence is taken over a particular class of
neighborhoods, called a \emph{differentiation basis}, and that basis supports
the \emph{Vitali property}. We define these concepts for the reader below
and use them to prove Theorem \ref{thm:discrete-pstate}.

Let $\pst{\mu}$ denote the restriction of $\mu$ to pasts, and let
$\pst{\mu}_{x_\ell\dots x_1}$ be the measure on pasts that precede the word
$w:= x_1\dots x_\ell$. These are given by:
\begin{align*}
    \Prob{\pst{\mu}}
    {x_0 \dots x_{-k}} &:= 
    \Prob{\mu}{x_{-k}\dots x_0}\\
    \Prob{\pst{\mu}_{x_\ell\dots x_1}}
    {x_0 \dots x_{-k}} &:= 
    \Prob{\mu}{x_{-k}\dots x_0 x_1 \dots x_\ell}
  ~.
\end{align*}
Then \cref{eq:discrete-pstate-def} can be recast in the form of a convergence of
likelihood ratios, taken over a sequence of cylinder sets $U_k:=U_{0,x_0 \dots
x_{-k}}$ converging on $\pst{x}$:
\begin{align}
\label{eq:discrete-pstate-deriv}
    \Prob{\mu}{x_1\dots x_\ell|\pst{X}} = 
    \lim_{k\rightarrow \infty}
    \frac{\pst{\mu}_{x_\ell\dots x_1}\left(U_k\right)}
    {\pst{\mu}\left(U_k\right)}
  ~.
\end{align}
This reformulation, though somewhat conceptually cumbersome, is useful due
to theorems that relate the convergence of likelihood ratios to the
Radon-Nikodym derivative. Indeed, wherever \cref{eq:discrete-pstate-deriv}
converges, it will be equal to the Radon-Nikodym derivative
$d\pst{\mu}_{x_\ell\dots x_1}/d\pst{\mu}(\pst{X})$.

\begin{figure*}[t]
\includegraphics[width=0.7\textwidth]{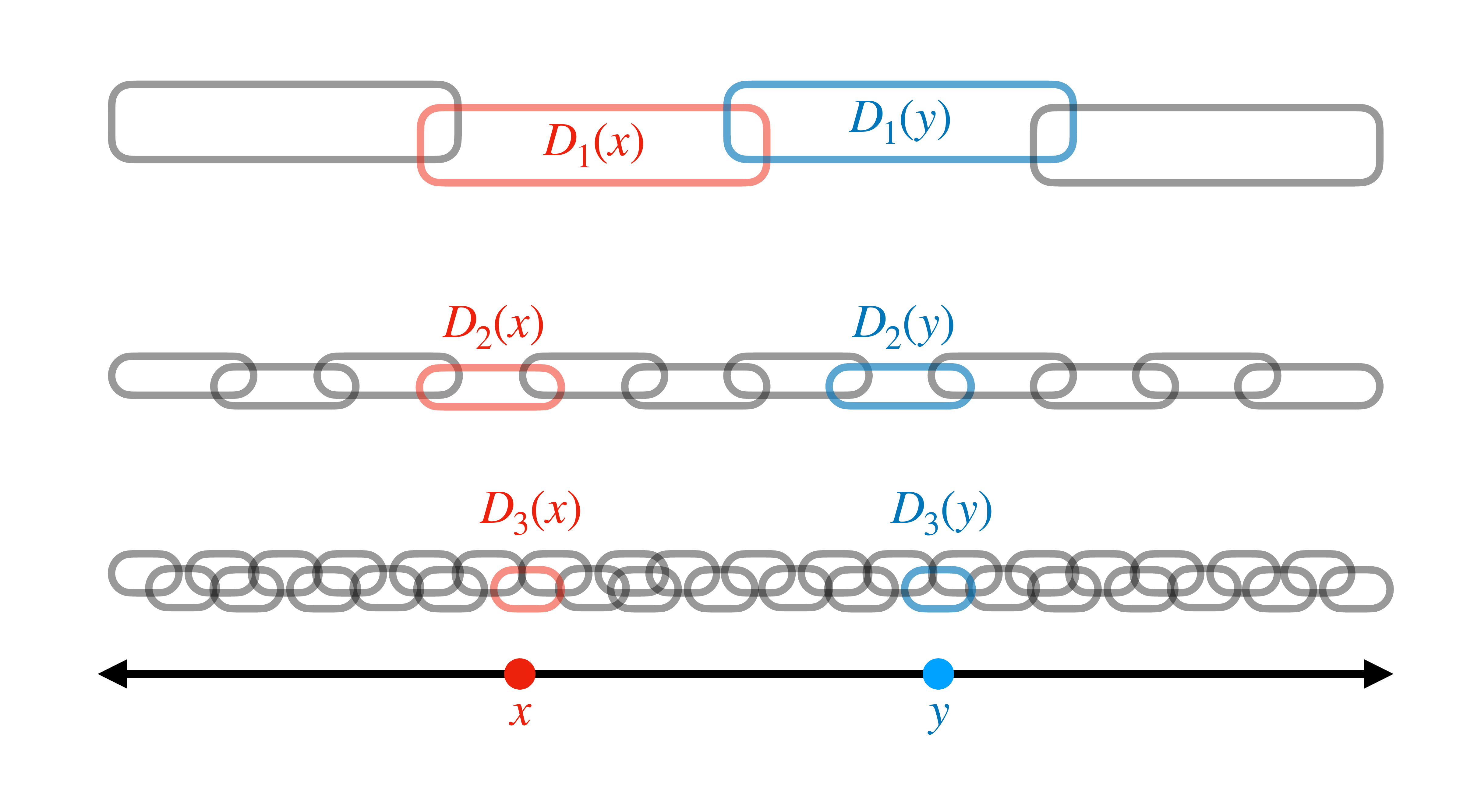}
\caption{\emph{Snapshot of a differentiation basis:} A differentiation basis is
	a collection of neighborhoods in $\mathcal{X}^\mathbb{N}$ that have
	hierarchical structure. For every point $x\in\mathcal{X}^\mathbb{N}$, there
	must be a sequence of neighborhoods converging on that point. Pictured
	above, a line is shown with a partial representation of its differentiation
	basis above it in the form of a hierarchical collection of rounded
	rectangles. For two points $x$ and $y$ we show the corresponding sequence
	of sets $(D_j(x))$, $(D_j(y))$ converging on each.
  }  
\label{fig:diff-basis}
\end{figure*}

To use these theorems we must define a \emph{differentiation basis}. Any
collection of neighborhoods $\mathcal{D}$ in $\mathcal{X}^\mathbb{N}$ may be
considered a differentiation basis if for every
$\pst{X}\in\mathcal{X}^\mathbb{N}$, there exists a sequence of neighborhoods $(D_k)$
such that $\lim_{k\rightarrow\infty}D_k = \Set{\pst{X}}$. 
See Fig. \ref{fig:diff-basis}.

The Vitali theorem states that whenever the differentiation basis $\mathcal{D}$
possesses the \emph{Vitali property} with respect to two measures $\nu$ and
$\mu$, then for $\mu$-almost all $\pst{X}$, the limit of likelihood ratios
exists for any sequence $(V_k)\subset \mathcal{D}$ converging on $\pst{X}$ and
the limit is equal to the Radon-Nikodym derivative $d\mu/d\nu(\pst{X})$ at that
point \cite{Rao05a}. This kind of very flexible limit is denoted by:
\begin{align*}
  \lim_{\substack{V\in\mathcal{D}\\ V\ni \pst{X}}}
  \frac{\mu(V)}{\nu(V)} = \frac{d\mu}{d\nu}(\pst{X})
  ~.
\end{align*}

The Vitali property has strong and weak forms, but we establish only the strong
form here. The differentiation basis $\mathcal{D}$ has the strong Vitali
property with respect to $\mu$ if for every measurable set $A$ and for every a
subdifferentiation basis $\mathcal{D}'\subseteq\mathcal{D}$ covering $A$, there
is an \emph{at most countable} subset $\Set{D_j}\subseteq \mathcal{D}'$ such
that $D_j\cap D_{j'}$ is empty for all $j\neq j'$ and:
\begin{align*}
    \mu\left(A-\left(\bigcup_j D_j\right)\right)
    = 0
  ~.
\end{align*}
In other words, we must be able to cover ``almost all'' of $A$ with a countable
number of nonoverlapping sets from the differentiation basis \cite{Rao05a}.

We now demonstrate that the differentiation basis $\mathcal{D}$ generated by
cylinder sets on $\mathcal{X}^\mathbb{N}$ has the Vitali property for \emph{any
measure} $\mu$.

\begin{Prop}[Vitali property for stochastic processes]
    For any stochastic process $(\mathcal{X}^\mathbb{N},\Sigma,\mu)$, let
    $\mathcal{D}$ be the differentiation basis of allowed cylinder sets. Then
    $\mathcal{D}$ has the strong Vitali property.
\end{Prop}

\begin{ProProp}
Let $\mathcal{D}'\subseteq\mathcal{D}$ be any subdifferentiation
basis covering $\mathcal{X}^\mathbb{N}$. (Our proof trivially generalizes to
any $A\subseteq \mathcal{X}^\mathbb{N}$.) Since $\mathcal{D}'$ is a
differentiation basis, for all $\pst{X}\in\mathcal{X}^\mathbb{N}$ there must be
a sequence $({D}_j(\pst{X}))$ of cylinder sets converging on $\pst{X}$.
Without loss of generality, suppose ${D}_j(\pst{X}) = U_{-\ell_j,
x_{-\ell_j+1}\dots x_0}$ with $\ell_j$ monotonically increasing. (If this is
not the case, we take a subsequence of ${D}_j(\pst{X})$ for which it is the
case.)

Now consider the combination of all such sequences:
\begin{align*}
    \mathcal{D}'' := \bigcup_{\pst{X}\in \mathcal{X}^\mathbb{N}}
    \Set{{D}_j(\pst{X}) | j\in\mathbb{N}} 
  ~.
\end{align*}
We note that $\mathcal{D}''$, though a union of an uncountable number of sets,
itself cannot be larger than a countable set, as the elements of the sets from
which it is composed are characterized by finite words, and finite words
themselves only form a countable set. That is, there is significant redundancy
in $\mathcal{D}''$ that keeps it countable. Furthermore, $\mathcal{D}''$ has a
lattice structure given by the set inclusion relation $\subseteq$ with the
particular property that for $U,V\in \mathcal{D}''$, $U\cap V$ is nonempty only
if $U\subseteq V$ or vice versa.

We then choose the set $\mathcal{C}$ of all maximal elements of this lattice:
that is, those $U\in \mathcal{D}''$ such that there is no $V\in \mathcal{D}''$
containing $U$. These maximal elements must exist since for each $U\in
\mathcal{D}''$ there is only a finite number of sets in $\mathcal{D}''$ that
can contain it.

It must be the case that all sets in $\mathcal{C}$ are nonoverlapping.
Furthermore, for any $V\in \mathcal{D}''$, not in $\mathcal{C}$, there
can only be a finite number of such sets containing $V$. One of them must be
maximal and therefore in $\mathcal{C}$. In particular, for every
$\pst{X}\in\mathcal{X}^\mathbb{N}$, each of its neighborhoods in $\mathcal{D}''$
is contained by the union of $\mathcal{C}$.

This implies $\mathcal{C}$ is a complete covering of $\mathcal{X}^\mathbb{N}$.
Since it is also nonoverlapping and countable, the strong Vitali property is
proven.
\end{ProProp}

As a consequence, the likelihood ratios in \cref{eq:discrete-pstate-deriv} must
converge for $\pst{\mu}$-almost every past $\pst{X}$ and every finite-length
word $w$---proving Theorem \ref{thm:discrete-pstate}.

We note that this result follows as a relatively straightforward application of
the Vitali property, which holds for any measure $\mu$ on
$\mathcal{X}^\mathbb{Z}$ and $\mathcal{X}^\mathbb{N}$. Our good fortune is due
to the particularly well-behaved topology of sequences of discrete observations.
For continuous observations, a less direct path to predictive states must be
taken.

\subsection{Continuous observations: Overview}
\label{sec:continuous}

Shifting from discrete to real-valued observations, where now $\mathcal{X}$
denotes a compact subset of $\mathbb{R}^d$, multiple subtleties come to the
fore. 

First, it must be noted that even in $\mathbb{R}$, the existence of a Vitali
property is not trivial. For the Lebesgue measure, only a weak Vitali property
holds, though this is still sufficient for the equivalence between
Radon-Nikodym derivatives and likelihood ratios. The differentiation basis in
this setting can be taken to be comprised of all intervals $(a,b)$ on the real
line.

Second, to go from $\mathbb{R}$ to $\mathbb{R}^d$, constraints must be placed
on the differentiation basis. An ``interval'' here is really the Cartesian
product of intervals, but for a Vitali property to hold we must only consider
products of intervals whose edges are held in a fixed ratio to one another, so
that the edges converge uniformly to zero. Likelihood ratios for fixed-aspect
boxes of this kind can converge to the Radon-Nikodym derivative \cite{Rao05a}.

This requirement poses a challenge for generalizing the Vitali property to
infinite dimensions, as we must to study sequences of real numbers. A
fixed-aspect ``box'' around a sequence of real numbers is not a practical
construction. In the empirical setting, we can only observe information about a
finite number of past outputs. We therefore cannot obtain any ``uniform''
knowledge of the entire past. That is, a direct generalization of the case for
$\mathbb{R}^d$ does not suffice.

However, integration and differentiation on infinite-dimensional spaces has been
considered before, mainly by Jessen \cite{Jess34a,Jess52a} and later Enomoto
\cite{Enom54a}. Their results focused on generalizing Lebesgue measure to
$(S^1)^\mathbb{N}$, where $S^1$ is the circle. This section shows that their
results can be significantly extended. The primary result we prove is a
generalization of Enomoto's Theorem \cite{Enom54a}:

\begin{The}[Generalized Enomoto's Theorem]
\label{thm:diff-uniform}
Let $\mathcal{X}$ be an interval of $\mathbb{R}$, and let $\mu$ be any
probability measure over $\mathcal{X}^\mathbb{N}$. Let
$f:\mathcal{X}^\mathbb{N}\rightarrow \mathbb{R}^{+}$ and let $F$ be its
indefinite integral under $\mu$. Let $\mathcal{V}$ denote the differentiation
basis consisting of sets of the form:
\begin{align*}
    V_{n,\delta}(\pst{X})
    = \Set{
        \pst{Y} | \left|y_j-x_j\right|<\delta,\ j=1,\dots,n
    }
	~ .
\end{align*}
Then:
\begin{align}
\label{eq:diff-uniform}
    \lim_{\substack{V\in\mathcal{V}\\
                        V\ni \pst{X}}} \frac{F(V)}{\mu(V)} = f(\pst{X})
  ~,
\end{align}
for $\pst{\mu}$-almost all $\pst{X}$.
\end{The}

Note that the resulting differentiation basis is a weaker form of that
considered above. Each $V_{n,\delta}$ is evidently a cylinder set, but of a
very particular kind. As we take $\delta\rightarrow 0$ and $n\rightarrow
\infty$, we extend the ``window'' of the cylinder set to the entire past while
simultaneously narrowing its width \emph{uniformly}. This turns out to be
sufficient to replicate the same effect as the fixed-aspect boxes in the
finite-dimensional case.

As a corollary of Theorem \ref{thm:diff-uniform}, we have the following result
for predictive states:

\begin{Cor}\label{thm:continuous-pstate} For all measures $\mu$ on
$\mathcal{X}^\mathbb{Z}$, where $\mathcal{X}\subset\mathbb{R}^d$ is a compact
set, all neighborhoods $U\subset\mathcal{X}^\ell$, and all $\ell\in\mathbb{N}$,
and for $\pst{\mu}$-almost all pasts $\pst{X}=x_0x_{-1}\dots$, the limit:
\begin{align}
\label{eq:continuous-pstate-def}
    \Prob{\mu}{U|\pst{X}} := 
    \lim_{n\rightarrow\infty}
    \frac{\mu(V_{n,\delta(n)}\times U)}{\mu(V)}
\end{align}
converges as long as $\delta(n)>0$ for all $n$ and $\delta(n)\rightarrow 0$.
\end{Cor}

Note here that we allowed $\mathcal{X}\subset \mathbb{R}^d$. This can be
obtained from Enomoto's theorem by simply reorganizing a sequence of
$d$-dimensional coordinates from $(\mathbf{x}_1,\mathbf{x}_2,\dots)$ to
$(x_{11},\dots,x_{d1},x_{12},\dots,x_{d2},\dots)$. Enomoto's theorem then
requires uniformity of the intervals across past instances as well as within
each copy of $\mathbb{R}^d$.

As before, the quantities $\Prob{\mu}{U|\pst{X}}$ define a unique measure
$\epsilon[\pst{X}]$ on $\mathcal{X}^\mathbb{N}$. It is determined by:
\begin{align*}
\epsilon[\pst{X}](U)=\Prob{\mu}{U|\pst{X}}
  ~.
\end{align*}

Enomoto's theorem itself is the capstone result in a sequence of theorems
initiated by Jessen \cite{Jess34a}. To prove Theorem \ref{thm:diff-uniform}, we
must start from the beginning, generalizing Jessen's results. Fortunately, the
bulk of the effort comes in generalizing the first of these results---Jessen's
correspondence principle. After this, the generalization follows quite
trivially from the subsequent theorems. The next section provides the full proof
for a generalized correspondence principle and explains how this result impacts
the proofs of the subsequent theorems. For completeness, we also give the full
proof of the generalized Enomoto's theorem, though it does not differ much from
Enomoto's---published in French---once the preceding theorems are secured.

\subsection{Jessen's correspondence principle}
\label{sec:jessen}

The Jessen and Enomoto theory rests on a profound correspondence between
cylinder sets on $\mathcal{X}^\mathbb{N}$ and intervals on $\mathbb{R}$. To
state it, we must define the concept of a net.

A net is similar to but formally separate from a differentiation basis, but like
the latter allows for a notion of differentiation, called
\emph{differentiation-by-nets}. This is weaker than the Vitali property on a
differentiation basis, but following on Jessen's work, Enomoto showed that
differentiation-by-nets can be extended to describe a particular differentiation
basis with the Vitali property.

\begin{figure}[t]
\includegraphics[width=0.9\columnwidth]{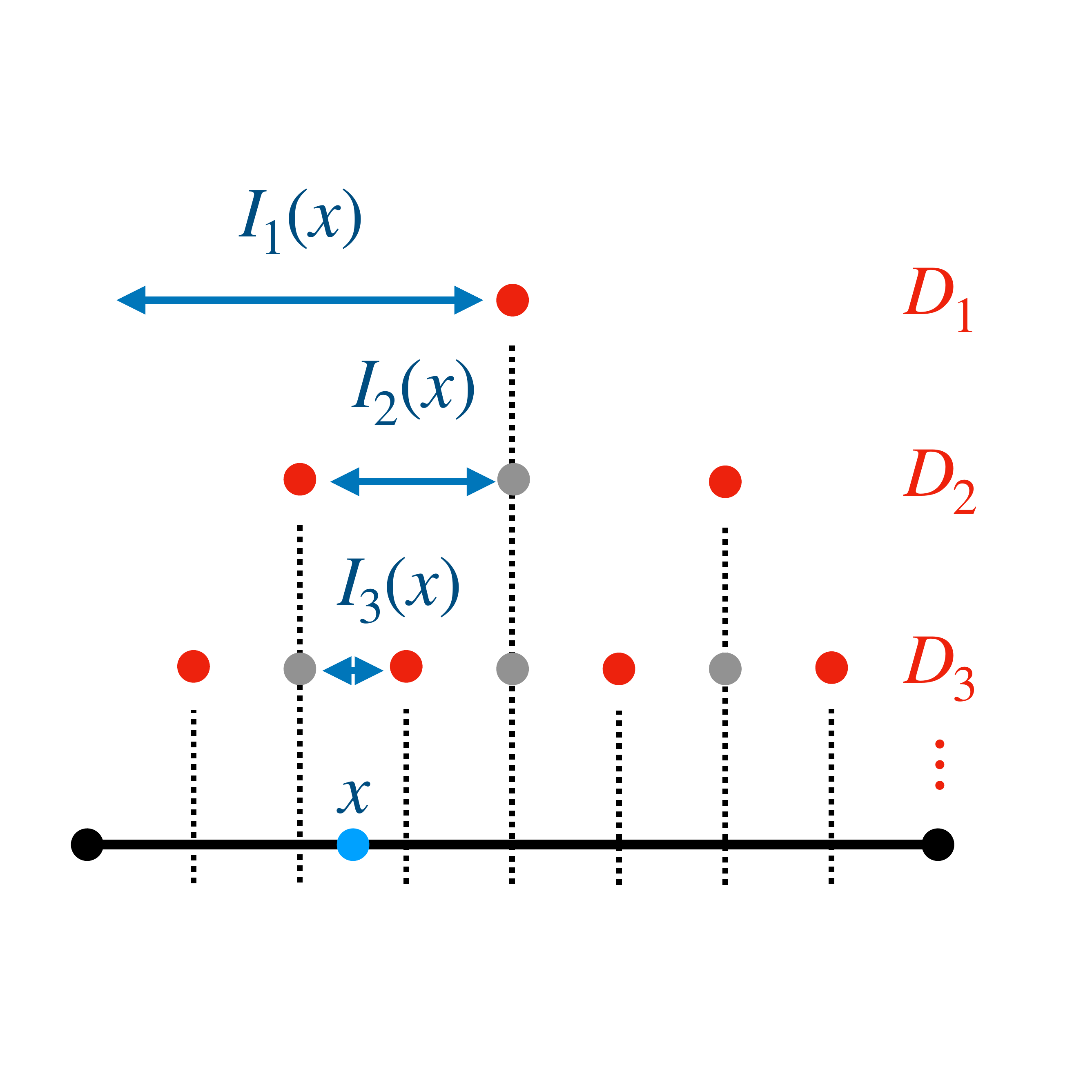}
\caption{\emph{Snapshot of a differentiation net.} 
  A differentiation net defined on a line segment. $D_1,D_2,D_3,\dots$
  represents the dissections which comprise the net. Each dissection contains
  the last; new points are indicated in red and old points in gray. These points
  define intervals; a sequence of these intervals is shown, $(I_k(x))$,
  converging on the point $x$.}  
  \label{fig:diff-net}
\end{figure}

Let $\mathcal{X}$ be a finite interval on $\mathbb{R}$. A \emph{dissection}
$D=(b_1,\dots,b_N)$ of $\mathcal{X}$ is simply a sequence of cut points, that
generate a sequence of adjacent intervals $(b_k,b_{k+1})$ spanning
$\mathcal{X}$, covering all but a finite set of points---the interval edges.
See Fig. \ref{fig:diff-net}. Denote the intervals
$\mathcal{I}(D)=\Set{(b_k,b_{k+1})|k=1,\dots,N-1}$. The length of the largest
interval in $\mathcal{I}(D)$ is denoted $|D|$. (Not to be confused with $D$'s
cardinality, that we have no need to reference.) A \emph{net} $\mathcal{N} =
(D_n)$ is a sequence of dissections so that $D_n\subset D_{n+1}$ (that is, each
new dissection only adds further cuts) and $|D_n|\rightarrow 0$ (the largest
interval length goes to zero). The boundary $\partial\mathcal{N} =
\bigcup_{n}D_n$ denotes all the boundary points from the sequence and is always
a countable set. 

\begin{figure*}[t]
  \includegraphics[width=0.9\textwidth]{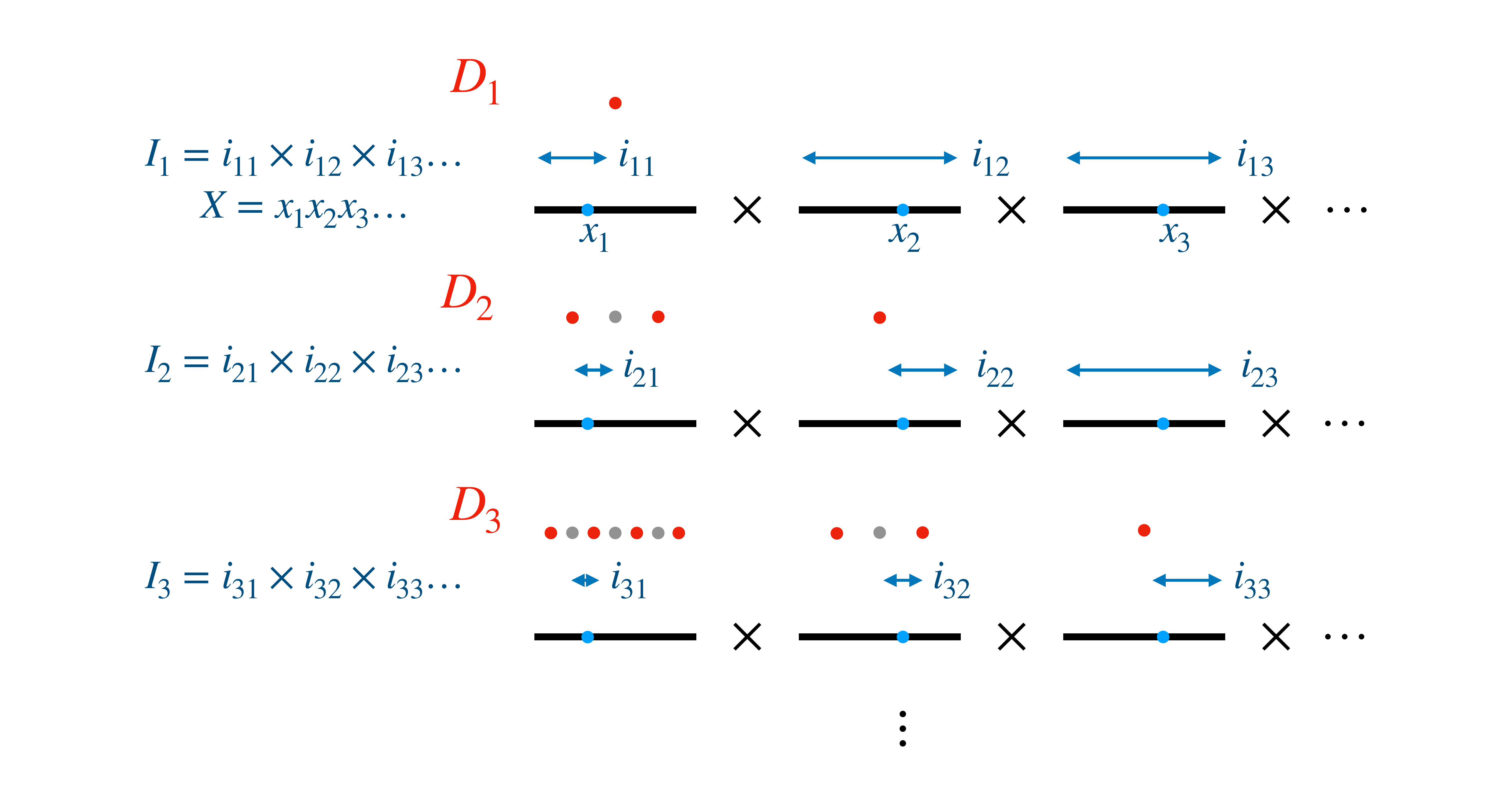}
  \caption{\emph{Snapshot of a differentiation net on a product space.} 
  A differentiation net defined on a product space $\mathcal{X}^\mathbb{N}$.
  This is comprised of an increasingly detailed dissection on each factor space.
  Also shown is a sequence of product intervals converging on a point
  $X=x_1x_2x_3\dots$.}  
  \label{fig:diff-net-product}
\end{figure*}

We can similarly define a dissection $D=(d_1,\dots,d_\ell)$ on
$\mathcal{X}^\mathbb{N}$ as a set of $\ell$ dissections, one for each of the
first $\ell$ copies of $\mathcal{X}$. $D$ intervals
$\mathcal{I}(D)=\Set{i_1\times\dots i_\ell\times \mathcal{X}^\mathbb{N} |
i_k\in\mathcal{I}(d_k)}$ are the cylinder sets generated by the intervals of
each individual dissection. See Fig. \ref{fig:diff-net-product}. The boundary
of a dissection is the set of all points that do not belong to these intervals:
$\partial D = \Set{X\in\mathcal{X}^\mathbb{N}| \exists k: x_k\in d_k}$. The
size of the dissection is $|D|:=\max_k |d_k|$. For a finite measure $\mu$,
there are always dissections with $\mu(\partial D)=0$ of any given $|D|=\max_k
|d_k|$, since $\mu|_{\mathcal{X}^\ell}$ can only have at most countably many
singular points.

A net $\mathcal{N}=\left(D_n=(d_{1,n},\dots,d_{\ell_n,n})\right)$ of
$\mathcal{X}^\mathbb{N}$ is a sequence of dissections of increasing depth
$\ell_n$ so that each sequence $(d_{k,n})$ for fixed $k$ is a net for the $k$th
copy of $\mathcal{X}$. $\partial\mathcal{N} = \bigcup_{n} \partial D_n$ denotes
all the accumulated boundary points of this sequence. Again, for finite measure
$\mu$, nets always exist that have $\mu(\partial\mathcal{N})=0$ for all $n$;
nets with this property are called \emph{$\mu$-continuous nets}.

Note that for any net, every sequence of intervals $(I_n)$,
$I_n\in\mathcal{I}(D_n)$ and $I_{n+1}\subset I_n$, uniquely determines a point
$X\in\mathcal{X}^\mathbb{N}$. If $X\not\in \partial D$, then $X$ uniquely
determines a sequence of intervals.

The following result can be proven (generalized from Ref. \cite{Jess34a}):
\begin{The}[Generalized correspondence principle]
\label{thm:correspondence}
Let $\mathcal{X}\subset \mathbb{R}$ be an interval and let $\lambda$ be the
Lebesgue measure on $\mathcal{X}$, normalized so $\lambda(\mathcal{X})=1$.  Let
$\mu$ be a finite measure on $\mathcal{X}^\mathbb{N}$ that has no singular
points. Let $\mathcal{N}=(D_n)$ be any $\mu$-continuous net of
$\mathcal{X}^\mathbb{N}$. Then there exists a net $\mathcal{M}=(d_n)$ of
$\mathcal{X}$ so that:
\begin{enumerate}
\item There exists a function $\Phi_n$ that maps each interval in
	$\mathcal{I}(D_n)$ of positive measure to one and only one interval in
    $\mathcal{I}(d_n)$ and vice versa for $\Phi_n^{-1}$;
\item $\lambda(\Phi_n(I)) = \mu(I)$ for all $I\in \mathcal{I}(D_n)$ with
    $\mu(I)>0$; and
\item The mapping $\phi:\mathcal{X}^\mathbb{N}-\partial\mathcal{N}
    \rightarrow \mathcal{X}-\partial\mathcal{M}$, generated by $X\mapsto
    (I_n) \mapsto (\Phi_n(I_n)) \mapsto x$, is measure-preserving.
\end{enumerate}
\end{The}

To summarize this technical statement: For any method of indefinitely
dissecting the set $\mathcal{X}^\mathbb{N}$ into smaller and smaller intervals,
there is in fact an ``equivalent'' such method for dissecting the much simpler
set $\mathcal{X}$. It is equivalent in the sense that all the resulting
intervals are in one-to-one correspondence with one another, a correspondence
that preserves measure. Since interval sequences uniquely determine points (and
vice versa for a set of full measure), this induces a one-to-one correspondence
between points that is also measure-preserving.

The proof consists of two parts. The first proves the first two claims about
$\mathcal{M}$. Namely, there is an interval correspondence and it is
measure-preserving. The second shows this extends to a correspondence between
$\mathcal{X}^\mathbb{N}$ and $\mathcal{X}$ that is also measure-preserving.

\begin{ProThe}[Interval correspondence]
The proof proceeds by induction. For a given $\mu$-continuous net
$\mathcal{N}=(D_n)$, suppose we already constructed dissections $d_1,\dots,d_N$
of $\mathcal{X}$ so that a function $\Phi_n$ between positive-measure intervals
in $D_n$ and $d_n$ exists with the desired properties (1) and (2) above, for all
$n=1,\dots,N$. Now, for $D_{n+1}$, a certain set of the intervals in
$\mathcal{I}(D_{n})$ is divided. Suppose $I\in \mathcal{I}_{n}$ divides
into $I'$ and $I''$. If either of these, say $I''$, has measure zero then we
discard it and set $\Phi_{n+1}(I')=\Phi_n(I)$. Otherwise, suppose that
$\Phi_n(I)=(a,b)$. Then divide $\Phi_n(I)$ into the intervals:
\begin{align*}
    \Phi_{n+1}(I') := \left(a,\frac{a\mu(I)+(b-a)\mu(I')}{\mu(I)}\right)\\
    \Phi_{n+1}(I'') := \left(\frac{a\mu(I)+(b-a)\mu(I')}{\mu(I)},b\right)
  ~,
\end{align*}
that clearly have Lebesgue measures $\lambda(\Phi_{n+1}(I'))=\mu(I')$ and
$\lambda(\Phi_{n+1}(I''))=\mu(I'')$, respectively. Generalizing this to more
complicated divisions of $I$ is straightforward.

Now, we can always suppose for a given net $\mathcal{N}$ that $D_0$ is just
the trivial dissection that makes no cuts and only one interval. However, this
has a trivial correspondence with $\mathcal{X}$; namely,
$\Phi_0(\mathcal{X}^\mathbb{N})=\mathcal{X}$.

By induction, then, the desired $\mathcal{M}$ can always be constructed.
\end{ProThe}

With the existence of the interval correspondence established, we further
demonstrate the existence of a point correspondence between $\mu$-almost-all of
$\mathcal{X}^\mathbb{N}$ and $\lambda$-almost-all of $\mathcal{X}$.

\begin{ProThe}[Point correspondence]
For every $X\in\mathcal{X}^\mathbb{N}-\partial \mathcal{N}$, there is a
unique sequence $(I_n)$ of concentric intervals, $I_n\in \mathcal{I}(D_n)$
and $I_{n+1}\subset I_{n}$, such that $\bigcap_n I_n = \{X\}$. If $X$ is in
the support of $\mu$, then we define:
\begin{align*}
    \phi(X) := \bigcap_{n}\Phi_n(I_n)
\end{align*}
as the corresponding point in $\mathcal{X}-\partial \mathcal{M}$. Due to the
interval correspondence, this mapping is invertible.
 
By measure-preserving we mean that for all $A\subseteq
\mathcal{X}^\mathbb{N}-\partial \mathcal{N}$, $\lambda(\phi(A))=\mu(A)$ and
vice-versa for $\phi^{-1}$. Both the Lebesgue measure and $\mu$ must be outer
regular, due to being finite measures. Outer regular means that the measure of
a set $A$ is the infimum of the measure of all open sets containing $A$, a
property we use to our advantage. 

Consider for each $n$ the minimal covering $\mathcal{C}_n$ of $A$ by intervals
in $\mathcal{I}(D_n)$. The measure of this covering is denoted $m_n :=
\mu(\bigcup\mathcal{C}_n)$. Clearly, $m_n\geq \mu(A)$ and $m_n\rightarrow
\mu(A)$. The corresponding covering $\Phi_n(\mathcal{C}_n)$ in
$\mathcal{I}(d_n)$ is a covering of $\phi(A)$ and has the same measure $m_n$.
By outer regularity, then, $m_n \geq \lambda(\phi(A))$ for all $n$. And so,
$\mu(A)\geq \lambda(\phi(A))$.

Now, by the exact reverse argument of the previous paragraph, going from
$\mathcal{X}$ to $\mathcal{X}^{\mathbb{N}}$ via $\phi^{-1}$, we can also deduce
that $\mu(A) \leq \lambda(\phi(A))$. Therefore $\mu(A) = \lambda(\phi(A))$, and
the function $\phi$ is measure-preserving.
\end{ProThe}

\subsection{Corollaries and Enomoto's Theorem}
\label{sec:enomoto}

Jessen's correspondence principle is an extremely powerful device. Among its
consequences are the following theorems regarding functions on
$\mathcal{X}^\mathbb{N}$. We state their generalized forms here and for the
proofs refer to Jessen \cite{Jess34a}, as each is a direct application of
Theorem \ref{thm:correspondence} without making any further assumptions on the
measure $\mu$.

The first offers a much weaker (and on its own, insufficient for our purposes)
concept of differentiation of measures that we refer to as
\emph{differentiation-by-nets}.

\begin{Cor}[Differentiation-by-nets]
\label{the:diff-nets}
	Let $f:\mathcal{X}^\mathbb{N}\rightarrow \mathbb{R}^{+}$ and let $F$ be the
	measure defined by its indefinite integral: $F(A):=\int_A f(X)d\mu(X)$.
	Further let $\mathcal{N}=(D_n)$ be a net on $\mathcal{X}^\mathbb{N}$ and
	denote by $\hat{f}_n$ a piecewise function such that
	$\hat{f}_n(X)=F(I_n)/\mu(I_n)$ for all $X\in I_n$ and each $I_n\in D_n$.
	Then $\hat{f}_n(X)\rightarrow f(X)$ as $n\rightarrow\infty$ for
	$\mu$-almost all $X$.
\end{Cor}

Though the full proof is found in Ref. \cite{Jess34a}, we summarize its key
point: Using the correspondence of intervals, we write
$F(I_n)/\mu(I_n)=\tilde{F}(\Phi(I_n))/\lambda(\Phi(I_n))$, where $\tilde{F}$ is
the indefinite integral of $f\circ \phi^{-1}$ with respect to $\lambda$. The
limit then holds due to the Vitali property of $\lambda$ on $\mathcal{X}$.
However, we also note that Corollary \ref{the:diff-nets} is \emph{not} an
extension of the Vitali property to cylinder sets on $\mathcal{X}^\mathbb{N}$.
Jessen himself offers a counterexample to this effect in a later publication
\cite{Jess52a}.

Jessen's second corollary is key to demonstrating that $\mathcal{V}$, the
differentiation basis defined in Theorem \ref{thm:diff-uniform}, \emph{will}
have the sought-after Vitali property. 

\begin{Cor}[Functions as limits of integrals]
\label{the:int-nets}
Let $f:\mathcal{X}^\mathbb{N}\rightarrow \mathbb{R}^{+}$, and let $f_n(X)$ be a
sequence of functions given by:
\begin{align*}
    f_n(x_1x_2\dots) := \int_{Y\in\mathcal{X}^\mathbb{N}}
        f(x_1\dots x_n Y) d\mu(Y)
  ~.
\end{align*}
That is, we integrated over all observations after the first $n$. Thus,
$f_n$ only depends on the first $n$ observations. Then $f_n(X)\rightarrow
f(X)$ as $n\rightarrow\infty$ for $\mu$-almost all $X$.
\end{Cor}

This proof we also skip, again referring the reader to Jessen \cite{Jess34a},
as no step is directly dependent on the measure $\mu$ itself and only on
properties already proven by the previous theorems.

We now have sufficient knowledge to prove the generalized Enomoto's theorem;
generalized from Ref. \cite{Enom54a}.

\begin{ProThe}[Generalized Enomoto's Theorem]
First, we must demonstrate, for almost every $X$, that there exists a sequence
$V_j(X)$ converging on $X$ such that the limit holds. By Corollary
\ref{the:int-nets}, there must be, for $\mu$-almost all $X$ and any
$\epsilon>0$, a $k(X,\epsilon)$ such that $|f_n(X)-f(X)|<\epsilon/2$ for all
$n>k(X,\epsilon)$. Now, from the Vitali property on $\mu_n$ and the fact that
$f_n$ only depends on the first $n$ observations, it must be true that for any
$\epsilon>0$ and almost all $X$, there is a $0<\Delta(X,n,\epsilon)<1$ so
that:
\begin{align*}
    \left|f_n(X) - \frac{F(V_{n,\delta}(X))}{\mu(V_{n,\delta}(X))}\right|
    <\epsilon/2
  ~,
\end{align*}
whenever $\delta<\Delta(X,n,\epsilon)$. For a given $\epsilon$, there is a
countable number of conditions (one for each $n$). As such, the set of points
$X$ for which all conditions hold is still measure one. Then, taking for each
$X$ the integer $K:=k(X,\epsilon)$ and subsequently the number $\Delta:=
\Delta(X,k(X,\epsilon),\epsilon)$, we can choose $V_{K,\Delta}(X)$ and by the
triangle inequality we must have:
\begin{equation}
   \left|f(X) - \frac{F(V_{K,\Delta}(X))}{\mu(V_{K,\Delta}(X))}\right|
   <\epsilon
  ~.
\end{equation}
This completes the proof's first part.

However, the second part---that \emph{all} sequences $V_{n_j,\delta_j}(x)$ of
neighborhoods give converging likelihood ratios---further follows from the
above statements, as:
\begin{align*}
    \left|f(X) - \frac{F(V_{n_j,\delta_j}(X))}{\mu(V_{n_j,\delta_j}(X))}\right|
    <\epsilon
\end{align*}
must hold for any $n_j>K(X,\epsilon)$ and any $\delta_j <
\Delta(X,K(X,\epsilon),\epsilon)$, which must eventually be true for any
converging sequence to $X$.
\end{ProThe}

Now, the previous theorem does not directly prove the Vitali property but rather
bypasses it. Demonstrating that the differentiation basis $\mathcal{V}$ may be
used to recover Radon-Nikodym derivatives. This, then, is sufficient for
Corollary \ref{thm:continuous-pstate} to hold, guaranteeing the existence of
predictive states $\epsilon[\pst{X}]$ for $\mu$-almost all $\pst{X}$.

\subsection{Remarks on convergence}
\label{sec:convergence}

An important task regarding predictive states is to learn a process's
predictive states---that is, the $\epsilon$-mapping from observed pasts to
distributions over futures---from a sufficiently large sample of observations.
These learned predictive states may then be used to more accurately predict the
process' future behavior based on behaviors already observed.

The previous three sections avoided constraining the measure $\mu$ on
$\mathcal{X}^\mathbb{N}$ to be anything other than finite. It was not assumed
to be either stationary or ergodic, in particular. In such cases the
$\epsilon$-mapping is time-dependent, as it obviously depends on where futures
are split from pasts. To adequately reconstruct $\epsilon[\pst{X}]$ from a
single, long observation requires that the process be both stationary and
ergodic.  Stationarity makes $\epsilon[\pst{X}]$ time-independent, and
ergodicity ensures that the probabilities in the limits
\cref{eq:discrete-pstate-def,eq:continuous-pstate-def} can be approximated by
taking the time-averaged frequencies of occurrence.

The next natural question is how rapidly convergence occurs for each past, in a
given process. So far, we only guaranteed that convergence exists, but said
nothing on its rate. This is process-dependent. Section \ref{sec:examples}
gives several examples of processes and process types with their convergence
rate. The most useful way to think of the rate is in the form of
``probably-almost-correct''-type statements, as exemplified in the following
result:

\begin{Prop}
Let $\mu$ be a probability measure on $\mathcal{X}^\ell$. Let
$\eta_{n,\delta}[\pst{X}](U)=\mu(V_{n,\delta}\times U)/\mu(V_{n,\delta})$.  For
every cylinder set $U$ and $\Delta_1,\Delta_2>0$, we have for sufficiently
large $\ell$ and small $\delta$:
\begin{align*}
    \Prob{\pst{\mu}}{\left|\eta_{\ell,\delta}[\pst{X}](U)-
    \epsilon[\pst{X}](U)\right|>\Delta_1}<\Delta_2
  ~.
\end{align*}
That is, the probability of an error beyond $\Delta_1$ is less than $\Delta_2$.
\end{Prop}
This is a consequence of the fact that all $\pst{X}$ must eventually converge.
The possible relationships between $\Delta_1$, $\Delta_2$, and $\ell$ in
particular is explored in our examples.

\section{Predictive states Form a Hilbert space}
\label{sec:kernel}

Thus far, we demonstrated that for discrete and real $\mathcal{X}$, measures
over $\mathcal{X}^\mathbb{N}$ possess a well-defined feature called
\emph{predictive states} that relate how past observations constrain future
possibilities. These states are defined by convergent limits that can be
approximated from empirical time series in the case of stationary, ergodic
processes.

We turn our attention now to the topological and geometric structure of these
states, the spaces they live in, and how the structure of these spaces may be
leveraged in the inference process. The results make contact between predictive
states \emph{as elements of a Hilbert space} and the well-developed arena of
reproducing kernel Hilbert spaces. To do this we introduce several new concepts.

Denote the set of real-valued continuous functions on $\mathcal{X}^\mathbb{Z}$ by $C(\mathcal{X}^\mathbb{Z})$. The set of signed measures on $\mathcal{X}^\mathbb{Z}$,
that we call $\mathbb{M}(\mathcal{X}^\mathbb{Z})$, may be thought of as
dual to $C(\mathcal{X}^\mathbb{Z})$. This allows us to define a notion of
convergence of measures on $\mathcal{X}^\mathbb{Z}$ in relation to continuous
functions. We say that a sequence of measures $\mu_n$ \emph{converges in
distribution} if:
\begin{align*}
   \lim_{n\rightarrow \infty} \int F(X)d\mu_n(X) = \int F(X)d\mu(X)
\end{align*}
for all $F \in C(\mathcal{X}^\mathbb{Z})$. Convergence in distribution is
sometimes referred to as weak convergence but we avoid this vocabulary to
minimize confusion---as another, distinct kind of weak convergence is needed in
the Hilbert space setting.

A kernel $k:\mathcal{X}^\mathbb{Z}\times \mathcal{X}^\mathbb{Z}\rightarrow
\mathbb{R}$ generates a \emph{reproducing kernel Hilbert space} (RKHS)
$\mathcal{H}$ if $k(\cdot,\cdot)$ is positive semi-definite and symmetric
\cite{Aron50a}. $\mathcal{H}$ is typically defined as a space of
\emph{functions} (from $\mathcal{X}^\mathbb{Z}\rightarrow \mathbb{R}$), but the
kernel allows embedding measures on $\mathcal{X}^\mathbb{Z}$ into the function
space through $f_\mu(x) = \int k(x,y)d\mu(y)$. This elicits an inner product
between any two positive measures $\mu$ and $\nu$:
\begin{align*}
    \left<f_\mu |f_\nu\right>_k := \int\int k(x,y)d\mu(x)d\nu(y)
  ~.
\end{align*}
The inner-product space on measures generated by this construction is isometric
to the RKHS generated by $k(\cdot,\cdot)$. The embedding of measures into this
space is unique if the kernel is \emph{characteristic}. And, convergence in the
norm of the Hilbert space is equivalent to convergence in distribution whenever
the kernel is \emph{universal} \cite{Srip10a}.

What exactly is the set $\mathcal{H}$ of functions? The equivalence of
convergence in norm and convergence in distribution tempts identifying
$\mathcal{H}$ with the space of continuous functions, but this is overly
optimistic. If it were true---that
$\mathcal{H}=C(\mathcal{X}^\mathbb{Z})$---then the convergence
$\braket{F|f_{\mu_n}}\rightarrow\braket{F|f_{\mu}}$ for every $F\in
\mathcal{H}$ implies $\mu_n\rightarrow\mu$ in norm. This, though, identifies
norm convergence on the Hilbert space with \emph{weak} convergence, which for
Hilbert spaces is identified as the convergence of every inner product.

For infinite-dimensional Hilbert spaces, these two types of convergence cannot
be identified. This is illustrated by the simple case of any orthogonal basis
$e_i$, for which $\braket{F|e_i}\rightarrow 0$ is necessary for $F$ to have a
finite norm, even though $\|e_i\|\rightarrow 1$ by definition. So, while
convergence in the norm of $\mathcal{H}$ is equivalent to convergence in
distribution of measures, we must conclude that $\mathcal{H}$ can only be a
proper subspace of the continuous functions; a fact also noted in
\cite{Stei20a}.

\subsection{Topology of predictive states}

Let the (closure of the) set of a process's predictive states be denoted by:
\begin{align*}
    \mathcal{K}(\mu)
	:= \overline{\Set{\epsilon[\pst{X}] | \pst{X}\in \mathcal{X}^\mathbb{N}}}
  ~.
\end{align*}
The closure is taken under convergence in distribution.

The relation between pasts and predictive states may be highly redundant. For
instance, in the process generated by the results of a random coin-toss, since
the future observations do not depend on past observations, $\mathcal{K}(\mu)$
is trivial. Meanwhile, for a periodic process of period $k$, $\mathcal{K}(\mu)$
has $k$ elements, corresponding to the $k$ distinct states---the process'
phases. In more complex cases, $\mathcal{K}(\mu)$ may have countable and
uncountable cardinality.

\begin{figure}[t]
\includegraphics[width=0.9\columnwidth]{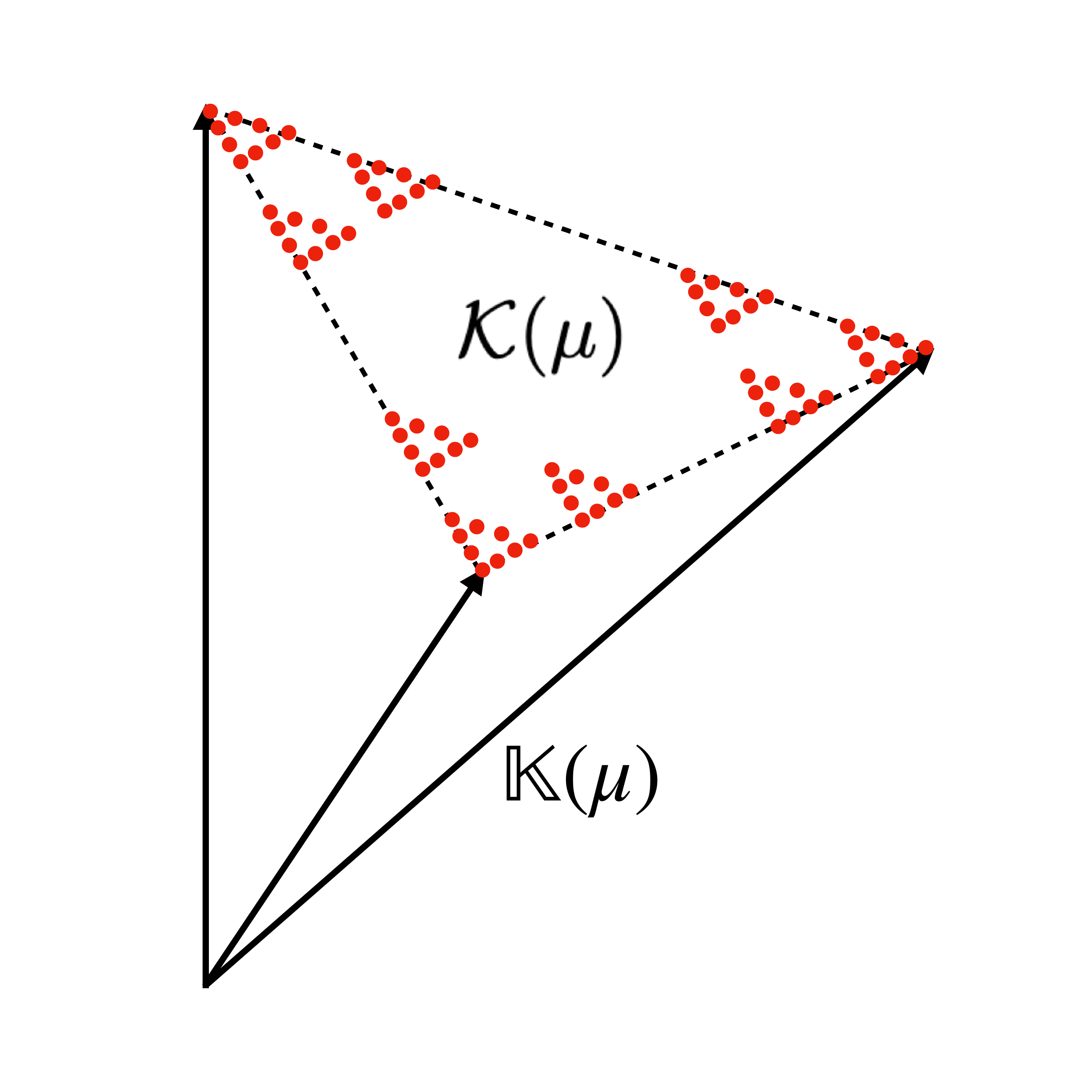}
\caption{\emph{Predictive states and their closed span.} The red dots are a
	hypothetical set $\mathcal{K}(\mu)$ of predictive states (shaped like the
	Sierpinski set) that is uncountably infinite but that has a
	finite-dimensional closed span $\mathbb{K}(\mu)$.
	}
\label{fig:predictive-span}
\end{figure}

We may also consider the vector space of signed measures generated by the
closed span of $\mathcal{K}(\mu)$, denoted $\mathbb{K}(\mu)$. This is the
smallest closed vector space that contains the predictive states. This, too,
may demonstrate redundancy in the form of linear dependence, regardless of the
cardinality of $\mathcal{K}(\mu)$. For instance, it is possible to have an
uncountably infinite set of predictive states $\mathcal{K}(\mu)$ whose
dimension, $\dim \mathbb{K}(\mu)$, is finite. In fact, it is the general case
that any process generated by an HMM or GHMM will have finite dimensional
$\mathbb{K}(\mu)$, but is not guaranteed to have finite $\mathcal{K}(\mu)$
\cite{Jurg20c}. See Fig. \ref{fig:predictive-span}.

The topology of convergence in distribution is closely related to the
definition of continuity on $\mathcal{X}^\mathbb{N}$. It behooves us at this
juncture to discuss $\mathcal{X}^\mathbb{N}$ not only as a topological space
but also as a metric space.

Two useful families of distance metrics, equivalent to the product topology on
$\mathcal{X}^\mathbb{N}$, are the Euclidean metrics, one for the discrete and
real case each:
\begin{align*}
    \begin{split}
    D{}_{\mathrm{E},\gamma}(X,Y)^2 
    := \begin{cases}
        \sum_{t=1}^\infty (1-\delta_{x_t y_t})\gamma^{2t} & \mathcal{X}\ \mathrm{discrete}\\
        \sum_{t=1}^\infty \left\|x_t-y_t\right\|^2\gamma^{2t}
        & \mathcal{X}\subset \mathbb{R}^d
    \end{cases}
\end{split}
  ~,
\end{align*}
for some $0<\gamma<1$. These distance metrics arise from embedding
$\mathcal{X}^\mathbb{N}$ in a Hilbert space. Given an orthogonal basis ($e_i$),
the components of this embedding for the discrete case are given by:
\begin{align*}
    c_i({X}) = \begin{cases}
        \gamma^{\lfloor i/|\mathcal{X}| \rfloor} 
        & x_{\lfloor i/|\mathcal{X}| \rfloor} = i \ \mathrm{mod}\ |\mathcal{X}|\\
        0 & \mathrm{otherwise}
    \end{cases}
\end{align*}
and in the continuous case ($\mathcal{X} \subset \mathbb{R}^d$) by:
\begin{align*}
    c_i({X}) = \gamma^{t} x_{k,t},\quad i=k \mod d 
  ~.
\end{align*}
Using these distance metrics, the following section introduces an inner product
structure on $\mathbb{K}(\mu)$ whose norm metrizes the topology of convergence
in distribution. Once $\mathbb{K}(\mu)$'s natural embedding into a Hilbert
space of its own is established, we investigate how well this embedding can be
approximated by the approach of reproducing kernel Hilbert spaces.

\subsection{Embedding predictions in a Hilbert space}
\label{sec:kernel-embed}

The space $\mathbb{K}(\mu)$ of predictive states is a subspace
$\mathbb{P}(\mathcal{X}^\mathbb{N})$ of the probability measures over
$\mathcal{X}^\mathbb{N}$. On $\mathbb{P}(\mathcal{X}^\mathbb{N})$, given any
symmetric positive-definite kernel
$k:\mathcal{X}^\mathbb{N}\times\mathcal{X}^\mathbb{N} \rightarrow \mathbb{R}$,
we can define an inner product over measures:
\begin{align}
\label{eq:inner-prod-def}
    \left<\mu,\nu\right>_k := \int\int k(X,Y)d\mu(X)d\nu(Y)
  ~.
\end{align}
Positive-definite means that for any finite set $\{X_i\}$ of
$X_i\in\mathcal{X}^\mathbb{N}$ and any set $\{c_i\}$ of values $c_i\in
\mathbb{R}$, both sets have the same cardinality:
\begin{align*}
    \sum_{i,j} k(X_i,X_j)c_i c_j \geq 0
  ~,
\end{align*}
with equality only when $c_i=0$ for all $i$. If this is true, then the inner
product \cref{eq:inner-prod-def} is positive-definite for all \emph{measures}.
That is, $\left<\mu,\mu\right>_k\geq 0$ with equality only when $\mu=0$
\cite{Srip10a}.

Since $\mathcal{X}^{\mathbb{N}}$ is compact, if the kernel $k$ satisfies the
property of being \emph{universal}, then norm convergence under the inner product
defined by $k$ is equivalent to convergence in distribution of measures
\cite{Srip10a}. A simple example of a universal kernel is the Gaussian radial
basis function, when paired with an appropriate distance---namely, one
defined from embedding $\mathcal{X}^\mathbb{N}$ in a Hilbert space, as our
$D_{\mathrm{E},\gamma}$ are \cite{Chri10a}. These take the form:
\begin{align*}
    k_{\beta,\gamma}(X,Y)
	:= \exp\left(-\frac{D_{\mathrm{E},\gamma}(X,Y)^2}{\beta^2}\right)
  ~.
\end{align*}
We denote the associated inner products by
$\left<\cdot,\cdot\right>_{\beta,\gamma}$. $\mathcal{H}_{\beta,\gamma} :=
\left(\mathbb{P}(\mathcal{X}^\mathbb{N}),
\left<\cdot,\cdot\right>_{\beta,\gamma}\right)$ defines a Hilbert space, since
it has the topology of convergence in distribution and
$\mathbb{P}(\mathcal{X}^\mathbb{N})$ is complete in this topology.

When referring to a measure $\mu$ as an element of $\mathcal{H}_{\beta,\gamma}$
we denote it $\ket{\mu}_{\beta,\gamma}$ and inner products in the bra-ket
are $\braket{\mu|\nu}_{\beta,\gamma}$. Now, it should be
noted that to every ket $\ket{\mu}_{\beta,\gamma}$ there is a bra
$\bra{\mu}_{\beta,\gamma}$ that denotes a dual element. However, the dual
elements of $\mathbb{P}(\mathcal{X}^\mathbb{N})$ correspond to continuous
functions. The function $f_\mu$ corresponding to $\bra{\mu}_{\beta,\gamma}$ is
given by:
\begin{align}
\label{eq:dual-function}
    f_\mu(X) := \int k_{\beta,\gamma}(X,Y)d\mu(Y)
  ~,
\end{align}
so that:
\begin{align*}
    \braket{\mu|\nu}_{\beta,\gamma} = \int f_\mu(X) d\nu(Y)
  ~.
\end{align*}
Let $\mathcal{F}_{\beta,\gamma}$ denote the space of all $f_\mu$ that can be
constructed from \cref{eq:dual-function}. This function space, when paired with
the inner product $\left<f_\mu,f_\nu\right>:=\braket{\nu|\mu}$, is isomorphic
to $\mathcal{H}_{\beta,\gamma}$. $\mathcal{F}_{\beta,\gamma}$ is then a
reproducing kernel Hilbert space with kernel $k_{\beta,\gamma}$.

As the start of Section \ref{sec:kernel} discussed, $\mathcal{F}_{\beta,\gamma}\subset
C(\mathcal{X}^\mathbb{N})$. Furthermore, the $\mathcal{F}_{\beta,\gamma}$ are
not identical to one another, obeying the relationship
$\mathcal{F}_{\beta,\gamma}\subset \mathcal{F}_{\beta',\gamma}$ when $\beta >
\beta'$ \cite{Zhan13a}. However, it is also the case that each
$\mathcal{F}_{\beta,\gamma}$ is dense in $C(\mathcal{X}^\mathbb{N})$, so their
representative capacity is still quite strong \cite{Srip10a}.

We note an important rule regarding the scaling of our inner products, as
constructed. The distances $D_{\mathrm{E},\gamma}(X,Y)$ have finite diameter on
our spaces. Let $\Delta$ denote the diameter of $\mathcal{X}$. For discrete
$\mathcal{X}$ we simply have $\Delta=1$; for $\mathcal{X}\subset\mathbb{R}^d$,
$\Delta$ is determined by the Euclidean distance. Then
$\mathcal{X}^\mathbb{N}$'s diameter is given by $\Delta/\sqrt{1-\gamma^2}$.
Since the Gaussian is bounded below by $1-D_\gamma^2/\beta^2$, for arbitrarily
large $\beta$:
\begin{align}
\label{eq:norm-bound}
    \left\|\mu-\nu\right\|_{\beta,\gamma}^2 \leq 
    \frac{\|\mu-\nu\|_{\mathrm{TV}} \Delta^2}{(1-\gamma^2)\beta^2} + O(\beta^{-3})
  ~,
\end{align}
where $\|\cdot\|_{\beta,\gamma}$ is simply the norm of
$\mathcal{H}_{\beta,\gamma}$ and $\|\cdot\|_{\mathrm{TV}}$ is the total
variation norm. This tells us that the norm is less discriminating between
measures as $\beta\rightarrow\infty$. Naturally, this can be remedied by
rescaling the kernel with a $\beta^2$ factor. As it happens,
\cref{eq:norm-bound} will be useful later.

\subsection{Finite-length embeddings}
\label{sec:kernel-truncate}

Our goal is to study how reproducing kernel Hilbert spaces may be used to encode
information about predictive states gleaned from empirical observations. Given
that such observations are always finite in length, we must determine
whether and in what manner the Hilbert space representations of measures over
finite-length observations converges to the Hilbert space representation of a
measure over infinite sequences.

Let $\mu_\ell$ denote the measure $\mu$ restricted to $\mathcal{X}^\ell$.
Define the restricted distance on $\mathcal{X}^\ell$:
\begin{align*}
\begin{split}
    D^{(\ell)}_{\mathrm{E},\gamma}(X,Y)^2 
    := \begin{cases}
    \sum_{t=1}^\ell (1-\delta_{x_t y_t})\gamma^{2t} & \mathcal{X}\ \mathrm{discrete}\\
    \sum_{t=1}^\ell \left\|x_t-y_t\right\|^2\gamma^{2t}
    & \mathcal{X}\subset \mathbb{R}^d
\end{cases}
\end{split}
  ~,
\end{align*}
for $X,Y\in\mathcal{X}^\ell$. This gives an important Pythagorean
theorem for sequences:
\begin{align}
\label{eq:pythag-distance}
\begin{split}
    D_{\mathrm{E},\gamma}(X,Y)^2
    =& D^{(\ell)}_{\mathrm{E},\gamma}(x_1\dots x_\ell,y_1\dots y_\ell)^2\\
    &+ \gamma^{2\ell}D_{\mathrm{E},\gamma}(x_{\ell+1}\dots,y_{\ell+1}\dots)^2
\end{split}
  ~.
\end{align}

Now, using $D^{(\ell)}_{\mathrm{E},\gamma}$ define kernels $k^{(\ell)}_{\beta,\gamma}$ in the same style as for $\mathcal{X}^\mathbb{N}$. These generate inner products on $\mathbb{P}(\mathcal{X}^\ell)$. Denote by $\mathcal{H}^{(\ell)}_{\beta,\gamma}$ the resulting Hilbert spaces. These are related to the original $\mathcal{H}_{\beta,\gamma}$ by the following
factorization theorem:

\begin{Prop}
    The predictive Hilbert space $\mathcal{H}_{\beta,\gamma}$ factors into
    $\mathcal{H}^{(\ell)}_{\beta,\gamma}\otimes
    \mathcal{H}_{\beta\gamma^{-\ell},\gamma}$.
\end{Prop}

Before stating the proof, we should explain the above. The factorization
$\mathcal{H}_{\beta,\gamma} =
    \mathcal{H}^{(\ell)}_{\beta,\gamma}\otimes
    \mathcal{H}_{\beta\gamma^{-\ell},\gamma}$
denotes a separation of the infinite-dimensional $\mathcal{H}_{\beta,\gamma}$
into two pieces---one of which is finite-dimensional, but retains the same
kernel parameters and another \emph{reparametrize} infinite-dimensional
Hilbert space. The reparametrization is $\beta\rightarrow
\beta\gamma^{-\ell}$. This constitutes, essentially, a renormalization-group
technique, in which the the topology of words starting at depth $\ell$ is
equivalent to a reparametrization of the usual topology. This
reparametrization works precisely due to the Pythagorean theorem for
sequences \cref{eq:pythag-distance}.

\begin{ProProp}
We are demonstrating an isomorphism---a particularly natural one. Let
$\delta_{X}$ be the Dirac delta measure concentrated on $X$. We note that for
any measure $\mu$:
\begin{align*}
    \ket{\mu}_{\beta,\gamma} = \int \ket{\delta_X}_{\beta,\gamma}d\mu(X)
	~.
\end{align*}
Now, consider the linear function from $\mathcal{H}_{\beta,\gamma}$ to
$\mathcal{H}^{(\ell)}_{\beta,\gamma}\otimes
\mathcal{H}_{\beta\gamma^{-\ell},\gamma}$ that maps:
\begin{equation*}
    \ket{\delta_X}_{\beta,\gamma} \mapsto 
    \ket{\delta_{x_1\dots x_\ell}}_{\beta,\gamma}^{(\ell)}\otimes
    \ket{\delta_{x_{\ell+1}\dots}}_{\beta\gamma^{-\ell},\gamma}
  ~,
\end{equation*}
for every $X$. Then by \cref{eq:pythag-distance} we can see that this preserves
the inner product and so is an isomorphism.
\end{ProProp}

Note that for any of these Hilbert spaces there exists an element corresponding
to the constant function $\boldsymbol{1}(X)=1$ for all $X$. This function
always exists in $\mathcal{F}_{\beta,\gamma}$. We denote its corresponding
element in $\mathcal{H}_{\beta,\gamma}$ as
$\bra{\boldsymbol{1}}_{\beta,\gamma}$, so that
$\braket{\boldsymbol{1}|\mu}_{\beta,\gamma}=1$ for all $\mu$. Then the operator
$\Pi^{(\ell)}_{\beta,\gamma}: \mathcal{H}_{\beta,\gamma}\rightarrow
\mathcal{H}^{(\ell)}_{\beta,\gamma}$ is given by:
\begin{align*}
    \Pi^{(\ell)}_{\beta,\gamma} := I^{(\ell)}\otimes 
    \bra{\boldsymbol{1}}_{\beta,\gamma}
  ~,
\end{align*}
where $I^{(\ell)}$ is the identity on $\mathcal{H}^{(\ell)}_{\beta,\gamma}$.
It provides the canonical mapping from a measure $\mu$ to its restriction
$\mu_\ell$: That is, $\Pi^{(\ell)}_{\beta,\gamma}\ket{\mu}_{\beta,\gamma} =
\ket{\mu_\ell}_{\beta,\gamma}^{(\ell)}$.

Consider the ``truncation error''---that is, the residual error remaining when
representing a measure by its truncated form $\mu_\ell$ rather than by its full
form $\mu$. We quantify this in terms of an embedding. That is, there exists an
embedding of truncated measures $\mathbb{P}(\mathcal{X}^\ell)$ into the space
of full measures $\mathbb{P}(\mathcal{X}^\mathbb{N})$ such that the distance
between any full measure and its truncated embedding is small:

\begin{The}
\label{thm:truncated-embedding}
There exist isometric embeddings $\mathcal{H}^{(\ell)}_{\beta,\gamma}\mapsto
\mathcal{H}^{(\ell')}_{\beta,\gamma}$ and $\mathcal{H}^{(\ell)}_{\beta,\gamma}
\mapsto \mathcal{H}_{\beta,\gamma}$ for any $\ell\leq \ell'$. Furthermore, let
$\mu$ be any measure and $\mu_\ell$ be its restriction to the first $\ell$
observations, and let $\ket{\hat{\mu}_\ell}_{\beta,\gamma}$ be the embedding of
$\mu_\ell$ into $\mathcal{H}_{\beta,\gamma}$. Then
$\ket{\hat{\mu}_\ell}_{\beta,\gamma}\rightarrow \ket{\mu}_{\beta,\gamma}$ as
$\ell\rightarrow\infty$, with $\|\mu-\hat{\mu}_\ell\|_{\beta,\gamma} \sim
O(\beta^{-1}\gamma^{\ell})$.
\end{The}

\begin{ProThe}
Let $\lambda_{\beta,\gamma}\in\mathbb{P}(\mathcal{X}^\mathbb{N})$ denote the
measure such that $\bra{\lambda_{\beta,\gamma}}_{\beta,\gamma} = \bra{\boldsymbol{1}}_{\beta,\gamma}$ for a given
$\beta,\gamma$. For a measure $\mu$ with restriction $\mu_\ell$ let
$\hat{\mu}_\ell$ denote the measure on $\mathcal{X}^\mathbb{N}$ with the
property:
\begin{equation*}
    \hat{\mu}_\ell(A\times B) = \mu_\ell(A)\lambda_{\beta\gamma^{-\ell},\gamma}(B)
  ~,
\end{equation*}
for $A\in \mathcal{X}^\ell$ and $B\in\mathcal{X}^\mathbb{N}$. Then the
mapping $\mu_\ell \mapsto \hat{\mu}_\ell$ is an isomorphism, since:
\begin{align*}
   \braket{\hat{\mu}_\ell | \hat{\nu}_\ell}_{\beta,\gamma}
   &= \int \int k_{\beta,\gamma}(X,Y) d\hat{\mu}_\ell(X) d \hat{\nu}_\ell(X)\\
   &= \int \int k_{\beta,\gamma}^{(\ell)}(x_1\dots x_\ell,y_1\dots y_{\ell}) d{\mu}_\ell d {\nu}_\ell
   \\
   & ~\times \int \int k_{\beta \gamma^{-\ell},\gamma}(x_\ell\dots,y_{\ell}\dots)
   d\lambda_{\beta\gamma^{-\ell},\gamma}d\lambda_{\beta\gamma^{-\ell},\gamma}\\
   &= \braket{\mu_\ell|\nu_\ell}_{\beta,\gamma}^{(\ell)}\int
   d\lambda_{\beta\gamma^{-\ell},\gamma} \\
   & = \braket{\mu_\ell|\nu_\ell}_{\beta,\gamma}^{(\ell)}
  ~.
\end{align*}

Now, as a result of \cref{eq:pythag-distance}, note that for any two measures $\mu$ and $\nu$: 
\begin{widetext}
\begin{align*}
    \left<\mu,\nu\right> = 
    \int d\mu_\ell(x_1\dots x_\ell)\int d\nu_\ell(y_1\dots y_\ell)
    \exp\left(-\beta^2 D^{(\ell)}_\gamma(x_1\dots x_\ell,y_1\dots y_\ell)\right)
    \braket{\mu(\cdot|x_1\dots x_\ell),\nu(\cdot|y_1\dots y_\ell)}_{\beta \gamma^\ell,\gamma}
\end{align*}
If we combine this fact with the bound \cref{eq:norm-bound}, we have the
result:
\begin{align*}
    \begin{split}
    \left\|\mu-\hat{\mu_\ell}\right\|_{\beta,\gamma}^2 &= 
    \int d\mu_\ell(x_1\dots x_\ell)\int d\mu_\ell(y_1\dots y_\ell)
    \exp\left(-\beta^2 D^{(\ell)}_\gamma(x_1\dots x_\ell,y_1\dots y_\ell)\right)
    \left\|\mu(\cdot|x_1\dots x_\ell)-\hat{\mu}_\ell(\cdot|y_1\dots y_\ell)
    \right\|_{\beta \gamma^{-\ell}}^2\\
    &\leq \int d\mu_\ell(x_1\dots x_\ell)\int d\mu_\ell(y_1\dots y_\ell)
    \frac{\|\mu-\hat{\mu}_\ell\|_{\mathrm{TV}} \Delta^2\gamma^{2\ell}}{(1-\gamma^2)\beta^2}
    = \frac{\|\mu-\hat{\mu}_\ell\|_{\mathrm{TV}} \Delta^2\gamma^{2\ell}}{(1-\gamma^2)\beta^2}
  ~.
	\end{split}
\end{align*}
Thus, $\|\mu-\hat{\mu}_\ell\|_{\beta,\gamma} \sim O(\beta^{-1}\gamma^{\ell})$.
\end{widetext}
\end{ProThe}

In summary, representing measures $\mu$ over $\mathcal{X}^\mathbb{N}$ by their
truncated forms $\mu_\ell$ leads to a Hilbert space representation that admits
an approximate isomorphism to the space of full measures. The resulting truncation error is of order $O(\beta^{-1}\gamma^{\ell})$. 

We close this part with a minor note about a lower bound on the distance
between measures. Given a word $w$, the function on $\mathcal{X}^\ell$ that
equals $1$ when $X=w$ and zero otherwise has a representation
$\ket{w}^{(\ell)}_{\beta,\gamma}$ in $\mathcal{H}^{(\ell)}_{\beta,\gamma}$. (This
follows since for finite $\mathcal{X}$, all functions on $\mathcal{X}^\ell$
belong to $\mathcal{F}_{\beta,\gamma}^{(\ell)}$.) The extension of this to
$\mathcal{H}_{\beta,\gamma}$ is $\ket{w}_{\beta,\gamma} :=
\ket{w}^{(\ell)}_{\beta,\gamma}\otimes
\ket{\lambda_{\beta,\gamma}}_{\beta\gamma^{-\ell},\gamma}$. This has the
convenient property that $\braket{w|\mu}_{\beta,\gamma}=\Prob{\mu}{w}$. Then,
by the Cauchy-Schwarz inequality, for any measures $\mu$ and $\nu$ and any word
$w$:
\begin{align}\label{eq:lower-bound}
    \|\mu-\nu\|_{\beta,\gamma} & \geq \frac{|\braket{w|\mu-\nu}|}
    {\sqrt{\braket{w|w}_{\beta,\gamma}}} \nonumber \\
    & = \frac{\left|\Prob{\mu}{w}-\Prob{\nu}{w}\right|}
    {\sqrt{\braket{w|w}_{\beta,\gamma}}}
  ~.
\end{align}
So, word probabilities function as lower bounds on the Hilbert space norm.

\subsection{Predictive states from kernel Bayes' rule}
\label{sec:kernel-bayes}

A prominent use of reproducing kernel Hilbert spaces is to approximate
empirical measures \cite{Muan17a}. Given a measure $\mu$ over a space
$\mathcal{X}$ and $N$ samples $X_k$ drawn from this space, one constructs an
approximate representation of $\mu$ via:
\begin{align*}
    \ket{\hat{\mu}} := \frac{1}{N}\sum_{k=1}^N \ket{\delta_{X_k}}
  ~.
\end{align*}
In other words, $\mu$ is approximated as a sum of delta functions centered
on the observations. Convergence of this approximation to $\ket{\mu}$ is
(almost surely) $O(N^{-1/2})$ \cite{Muan17a}.

This fact, combined with our Theorem \ref{thm:truncated-embedding}, immediately
gives the following result for $\mathcal{H}_{\beta,\gamma}$:

\begin{Prop}
Suppose for some $\mu\in\mathbb{P}(\mathcal{X}^\mathbb{N})$ we take $N$ samples
of length $\ell$, denoted $\{X_k\in\mathcal{X}^\ell\}$ ($k=1\dots N$), and
construct the state:
\begin{align*}
    \ket{\hat{\mu}_{\ell,N}}_{\beta,\gamma} = \frac{1}{N}\sum_{k=1}^N 
    \ket{\delta_{X_k}}_{\beta,\ell}^{(\ell)}\otimes 
    \ket{\lambda_{\beta\gamma^{-\ell},\gamma}}_{\beta\gamma^{-\ell},\gamma}
  ~.
\end{align*}
Then $\ket{\hat{\mu}_{\ell,N}}_{\beta,\gamma} \rightarrow \ket{\mu}$ converges
almost surely as $N,\ell\rightarrow \infty$ with error
$O(N^{-1/2}+\beta^{-1}\gamma^\ell)$.
\end{Prop}

A more nuanced application of RKHS for measures lies in reconstructing
conditional distributions \cite{Song09a,Song10a,Fuku13a,Boot13a,Muan17a}. Let
$\mu$ be a joint measure on some $\mathcal{X}\times\mathcal{Y}$, and let
$\mu|_\mathcal{X}$ and $\mu|_{\mathcal{Y}}$ be its marginalizations. Given $N$
samples $(X_k,Y_k)$, construct the covariance operators:
\begin{align*}
    \hat{C}_{XX} & := \frac{1}{N}\sum_k \ket{\delta_{X_k}}\bra{\delta_{X_k}}
	~\text{and} \\
    \hat{C}_{YX} & := \frac{1}{N}\sum_k \ket{\delta_{Y_k}}\bra{\delta_{X_k}}
  ~.
\end{align*}
Let $\mu_{\mathcal{Y}|X}$ be the conditional measure for $X\in\mathcal{X}$.
For some $g\in\mathcal{H}_{\mathcal{Y}}$---the RKHS constructed on
$\mathcal{Y}$---let $F_g(X) := \braket{g|\mu_{\mathcal{Y}|X}}$ be a function on
$\mathcal{X}$. If $F_g\in \mathcal{H}_{\mathcal{X}}$ for all
$g\in\mathcal{H}_{\mathcal{Y}}$, then $\hat{C}_{YX}\left(\hat{C}_{XX}-\zeta
I\right)^{-1}\ket{\delta_X}$ converges to $\ket{\mu_{\mathcal{Y}|X}}$ as
$N\rightarrow\infty$, $\zeta\rightarrow 0$, with convergence rate
$O\left((N\zeta)^{-1/2}+\zeta^{1/2}\right)$.

The requirement essentially tells us that the structure of the conditional measure is compatible with the structures represented by the RKHS.

This is the \emph{kernel Bayes' Rule} \cite{Fuku13a}. It applies to our
$\mathcal{H}_{\beta,\gamma}$, by combining it with our results on truncated
representations:

\begin{The}\label{thm:kernel-bayes-rule}
Let $\mu\in\mathbb{P}(\mathcal{X}^\mathbb{Z})$ be a stationary and ergodic
process. Suppose we take a long sample $X\in \mathcal{X}^L$ and from this
sample subwords of length $2\ell$, $w_t=x_{t-\ell+1}\dots w_{t+\ell}$ for
$t=\ell,\dots,L-\ell$. (There are $L-2\ell+1$ such words.) Split each word
into a past $\pst{w}_t=x_{t-\ell+1}\dots w_t$ and a future $\fut{w}_t =
x_{t+1}\dots x_{t+\ell}$, each of length $\ell$. Define the operators:
\begin{align*}
  \hat{C}^{(\pst{X}\pst{X})}_{\beta,\gamma}
  & = \frac{1}{L-2\ell+1}\sum_{t=\ell}^{L-\ell}
  \ket{\hat{\delta}_{\pst{w}_t}}_{\beta,\gamma}\otimes 
  \ket{\hat{\delta}_{\pst{w}_t}}_{\beta,\gamma}
  ~\text{and} \\
  \hat{C}^{(\pst{X}\fut{X})}_{\beta,\gamma}
  & = \frac{1}{L-2\ell+1}\sum_{t=\ell}^{L-\ell}
  \ket{\hat{\delta}_{\pst{w}_t}}_{\beta,\gamma}\otimes 
  \ket{\hat{\delta}_{\fut{w}_t}}_{\beta,\gamma}
  ~.
\end{align*}
Now, suppose for every $g\in \mathcal{F}_{\beta,\gamma}$ that
$\left<\epsilon[\pst{X}],g\right>\in\mathcal{F}_{\beta,\gamma}$ and
$\left<\eta_\ell[\pst{X}],g\right>\rightarrow
\left<\epsilon[\pst{X}],g\right>$ at a rate of $O(h_{\pst{X}}(\ell))$;
see Section \ref{sec:convergence}.
Then for all $\pst{X}$:
\begin{align*}
\hat{C}^{(\pst{X}\fut{X})}_{\beta,\gamma}
    \left(\hat{C}^{(\pst{X}\pst{X})}_{\beta,\gamma}-\zeta\cdot
    I_{\beta,\gamma}\right)^{-1} \ket{\delta_{\pst{X}}}_{\beta,\gamma}
\end{align*}
almost surely converges to $\ket{\epsilon[\pst{X}]}_{\beta,\gamma}$ as
$L\rightarrow \infty$, $\ell\rightarrow \infty$, and $\zeta\rightarrow 0$, at
the rate $O\left( (L\zeta)^{-1/2} + \zeta^{1/2} + \gamma^{-\ell}
+h_{\pst{X}}(\ell)\right)$.
\end{The}

This integrates all our results thus far with the usual kernel Bayes' rule.
Several observations are in order. First, there will ($\pst{\mu}$-almost)
always be an $h_{\pst{X}}(\ell)$ as required by this theorem due to our own
Thm. \ref{thm:discrete-pstate} and Cor. \ref{thm:continuous-pstate}. Second,
since $\epsilon[\pst{X}]$ is not generally continuous, the theorem's
strict requirements on $\epsilon[\pst{X}]$ are not satisfied. That said, weaker
versions hold. If $\left<\epsilon[\pst{X}],g\right>$ as a function of $\pst{X}$
does not belong to $\mathcal{F}_{\beta,\gamma}$ as a function of $\pst{X}$,
then the representational error scaling depends on the precise form of
$\epsilon[\pst{X}]$. The latter can be obtained by choosing the
$\zeta$-parameter through cross-validation analysis \cite{Fuku13a,Muan17a}.

\section{Examples}
\label{sec:examples}

We close with a handful of examples and case studies that give further insight
to the convergence of $\big\| \eta_{\ell}[\pst{X}]-\epsilon[\pst{X}]
\big\|_{\beta,\gamma}$ for widely-employed process classes---Markov, hidden
Markov, and renewal processes.

\subsection{Order-$R$ Markov processes}
\label{sec:markov}

A \emph{Markov process} is a stochastic process where each observation $x_t$
statistically depends only on the previous observation $x_{t-1}$. An order-$R$
Markov process is one where each observation $x_t$ depends only on the previous
$R$ observations $x_{t-R}\dots x_{t-1}$. As such, the predictive states are
simply given by:
\begin{align*}
    \Prob{\mu}{x | \pst{X}} = \frac{\Prob{\mu}{x_{-R+1}\dots x_0 x}}
    {\Prob{\mu}{x_{-R+1}\dots x_0}}
  ~,
\end{align*}
for each $\pst{X} = x_0 x_{-1}\dots$. Since the predictive state is entirely
defined after a finite number of observations, and this number is bounded by
$R$, there is no conditioning error when $R$ is taken as the observation length.

\subsection{Hidden Markov processes}
\label{sec:hmms}

A \emph{hidden Markov model} (HMM)
$(\mathcal{S},\mathcal{X},\Set{\mathbf{T}^{(x)}})$ is defined here as a finite
set $\mathcal{S}$ of states, a set $\mathcal{X}$ of observations, and a set
$\mathbf{T}^{(x)}=({T}^{(x)}_{ss'})$ of transition matrices, labeled by
elements $x \in \mathcal{X}$ and whose components are indexed by $\mathcal{S}$
\cite{Uppe97a}. The elements are constrained so that $0 \leq {T}^{(x)}_{ss'} \leq 1$ and
$\sum_{x,s'}{T}^{(x)}_{ss'}=1$ for all $s, s' \in \mathcal{S}$. Let $\mathbf{T} =
\sum_x \mathbf{T}^{(x)}$ and $\boldsymbol{\pi}$ be its left-eigenvector such
that $\boldsymbol{\pi}\mathbf{T}=\boldsymbol{\pi}$. HMMs generate a stochastic
process $\mu$ defined by the word probabilities:
\begin{align*}
    \Prob{\mu}{x_1\dots x_\ell} := \sum_{s'}\left[\boldsymbol{\pi} \mathbf{T}^{(x_1)}\dots 
    \mathbf{T}^{(x_\ell)}\right]_{s'}
  ~.
\end{align*}

An extension of HMMs, called \emph{generalized hidden Markov models} (GHMMs)
\cite{Uppe97a} (or elsewhere \emph{observable operator models} \cite{Jaeg00a}),
is defined as $(\mathbf{V},\mathcal{X},\Set{\mathbf{T}^{(x)}})$ where
$\mathbf{V}$ is a finite-dimensional vector space. The only constraint on the
transition matrices $\mathbf{T}^{(x)}$ is that $\mathbf{T}$ have a simple
eigenvector of eigenvalue $1$, the left-eigenvector is still denoted
$\boldsymbol{\pi}$, the right-eigenvector denoted $\boldsymbol{\phi}$, and the
word probabilities:
\begin{align*}
   \Prob{\mu}{x_1\dots x_\ell} := \boldsymbol{\pi} \mathbf{T}^{(x_1)}\dots 
   \mathbf{T}^{(x_\ell)}\boldsymbol{\phi}
\end{align*}
are positive \cite{Uppe97a}. GHMMs generate a strictly broader class of
processes than finite hidden Markov models can \cite{Ito92a,Uppe97a,Jaeg00a},
though their basic structure is very similar.

First off, consider sofic processes. A \emph{sofic process} is one that is not
Markov at any finite order, but that is still expressible in a certain finite
way. Namely, a sofic process is any that can be generated by a finite-state
hidden Markov model with the \emph{unifilar property}. An HMM has the unifilar
property if $T^{(x)}_{s's}>0$ only when $s'=f(x,s)$ for some deterministic
function $f:\mathcal{S}\times\mathcal{X}\rightarrow \mathcal{S}$. Unifilar HMMs
are the stochastic generalization of deterministic finite automata in
computation theory \cite{Hopc06a}.

The most useful property of sofic processes is that the states of their minimal
unifilar HMM correspond exactly to the predictive states, of which there is
always a finite number. Unlike with order-$R$ Markov processes, there is no
upper bound to how many observations it may take to $\delta$-synchronize the
predictive states. However, closed-form results on the synchronization to
predictive states for unifilar HMMs is already known: at $L$ past observations,
with $L\rightarrow \infty$, the conditioning error is exponentially likely (in
$L$) to be exponentially small (in $L$) \cite{Trav10b}. In terms of our
Hilbert space norm, there are constants $\alpha$ and $C$ such that
\begin{align*}
    \Prob{\pst{\mu}}{\left\Vert\eta_{x_1\dots x_\ell}-
    \epsilon[\pst{X}]\right\Vert_{\beta,\gamma}>\alpha^\ell}
    < C \alpha^\ell
  ~.
\end{align*}
As such, for $\pst{\mu}$-almost-all pasts, the corresponding convergence rate
for the kernel Bayes' rule applied to a sofic process is
$O\left((L\zeta)^{-1/2}+\zeta^{1/2}+\min(\alpha,\gamma)^{-\ell}\right)$.

Not all discrete-observation stochastic processes can be generated with a
finite-state unifilar hidden Markov model. Though still encompassing only a
small slice of processes, generalized hidden Markov models have a considerably
larger scope of representation than finite unifilar models, as noted above.

The primary challenge in this setting is to relate the structure of a given HMM
to the predictive states of its process. This is achieved through the notion of
mixed states. A \emph{mixed state} $\rho$ is a distribution over the states of
a finite HMM. A given HMM, with the stochastic dynamics between its own states,
induces a higher-order dynamic on its mixed states and, critically for
analysis, this is an \emph{iterated function system} (IFS). Under suitable
conditions the IFS has a unique invariant measure, and the support of this
measure maps surjectively onto the process' set of predictive states. See Refs.
\cite{Jurg20c} for details on this construction.

If $\rho = (\rho)$ is a mixed state, then the updated mixed state after
observing symbol $x$ is:
\begin{align*}
  f^{(x)}_s(\rho) := \frac{1}
  {\sum_{s'}\left[\mathbf{T}^{(x)}\boldsymbol{\rho}\right]_{s'}}
  \left[\mathbf{T}^{(x)}\boldsymbol{\rho}\right]_{s}
  ~.
\end{align*}
Let the matrix $\left[\mathbf{D}f^{(x)}\right]_{s's}(\rho)$ be given by the
Jacobian $\partial f^{(x)}_{s'}/\partial \rho_s$ at a given value of $\rho$.
There is a statistic, called the \emph{Lyapunov characteristic exponent}
$\lambda < 0$, such that:
\begin{align*}
  \lambda = \lim_{\ell\rightarrow\infty}\frac{1}{\ell}
  \log\frac{\left\|\mathbf{D}f^{(x_\ell)}(\rho_\ell)
  \cdots \mathbf{D}f^{(x_1)}(\rho_1)\mathbf{v}\right\|}
{\left\|\mathbf{v}\right\|}
  ~,
\end{align*}
where $\rho_t := f^{(x_{t-1})}\circ\cdots\circ f^{(x_1)}(\rho)$, for any vector
$\mathbf{v}$ tangent to the simplex, almost any $\rho$ (in the invariant
measure), and almost any $\fut{X}=x_1x_2\dots$ (in the measure of the prediction
induced by $\rho$). The exponent $\lambda$ then determines the rate at which
conditioning error for predictive states converges to zero: for all
$\epsilon$ and sufficiently large $\ell$:
\begin{align*}
    \Prob{\pst{\mu}}{\left\Vert\eta_{x_1\dots x_\ell}-\epsilon[\pst{X}]\right\Vert_{\beta,\gamma}
    < C e^{\lambda \ell}}
    > 1-\epsilon
  ~.
\end{align*}
This is somewhat less strict---depending on how rapidly the Lyapunov exponent
converges in probability. In any case, for $\pst{\mu}$-almost all pasts, the
convergence of the kernel Bayes' rule is $O\left((L\delta)^{-1/2} +
\delta^{1/2}+\min(\lambda,\gamma)^{-\ell}\right)$, very similar to the sofic
process rate.

We anticipate that these rules still broadly apply to generalized hidden Markov
models, though we recommend more detailed analysis on this question.

\subsection{Renewal processes}
\label{sec:renewal}

A renewal process, usually defined over continuous-time, can be defined for
discrete time as follows. A renewal process emits $0$s for a randomly selected
duration before emitting a single $1$ and then randomly selecting a new
duration to fill with $0$s \cite{Marz17b}. Renewal processes can be as simple
as Poisson processes, where the probability at any moment of producing another
$0$ or restarting on $1$ is independent of time. Or, they can be far more
memoryful, with a unique predictive state for any number of past $0$s.

While high-memory renewal processes cannot generally be represented by a finite
hidden Markov model, they have only a countable number of predictive states,
unlike most hidden Markov models. This follows since every number of past $0$s
defines a potential predictive state, but the process has no further memory
beyond the most recent $1$. Said simply, the predictive states are the time
since last $1$---or some coarse-graining of this indicator in special cases,
such as the Poisson process.

A \emph{renewal process} is specified by the survival probability $\Phi(n)$
that a contiguous block of $0$s has length at least $n$. The exact probability
of a given length is $F(n) := \Phi(n)-\Phi(n+1)$. It is always assumed that
$\Phi(1)=1$. Further, stationarity requires that $m:=\sum_{n=1}^\infty\Phi(n)$
be finite, as this gives the mean length of a block of $0$s. In the most
general case the predictive states are given by:
\begin{align*}
\epsilon[\pst{X}]
    = \begin{cases}
        \epsilon_k & \pst{X}=0^{k}1 \dots \\
        \mathrm{undefined} & \pst{X}=0^\infty
    \end{cases}
  ~,
\end{align*}
where the measures $\epsilon_k$ are recursively defined by the word
probabilities:
\begin{align*}
    \Prob{\epsilon_k}{0^\ell 1w}
    = \frac{F(k+\ell)}{\Phi(k)}\Prob{\epsilon_0}{w}
  	~.
\end{align*}

Now, it can be easily seen that each past $\pst{X}$ converges to zero
conditioning error at a finite length since (almost) all pasts have the
structure $\dots 1 0^k$, and so only the most recent $k+1$ values need be
observed to know the predictive state. Therefore the kernel Bayes' rule has an
asymptotic convergence rate for each past $\pst{X}$ of
$O\left((L\delta)^{-1/2}+\delta^{1/2}+\gamma^{-\ell}\right)$. However, this
does not tell the entire story, as obviously not all pasts converge uniformly.
A probabilistic expression of the conditioning error gives more information:

\begin{Prop}
Suppose $\mu$ is a renewal process with $\Phi(n)\propto n^{-\alpha}$,
$\alpha>1$. Then there exist constants $C$ and $K$ such that:
\begin{align*}
   \Prob{\pst{\mu}}{
   \left\Vert \eta_{x_1\dots x_\ell}-\epsilon[\pst{X}] \right\Vert_{\beta,\gamma} >C\ell^{-1}
   } > K\ell^{-\alpha}
  ~.
\end{align*}
That is, the probability the conditioning error decays as $1/\ell$ is itself at
least power-law decaying in $\ell$.
\end{Prop}

\begin{ProProp}
Recall from \cref{eq:lower-bound}:
\begin{align*}
   &\left\| \eta_{x_1\dots x_\ell}-\epsilon[\pst{X}] \right\|_{\beta,\gamma} \\
        &\quad > 
   \frac{\left|\Prob{\mu}{w|x_1\dots x_\ell}-\Prob{\mu}{w|\pst{X}}\right|}
   {\sqrt{\braket{w|w}_{\beta,\gamma}}}
  ~,
\end{align*}
for every word $w$, so we can choose any $w$ and obtain a lower bound on the
conditioning error. If our past $\pst{X}$ has the form $0^k 1 \dots$ for
$k<\ell$, then we are already synchronized to the predictive state and the
conditioning error is zero. Thus, we are specifically interested in the case
$k\geq \ell$ and we will further consider the large-$\ell$ limit.

Now, under our assumptions, $\Phi(n)= n^{-\alpha}$ for some constant $Z$. For
large $n$, $F(n) \sim \alpha n^{-\alpha-1}$. Then for any $j$:
\begin{align*}
\Prob{\mu}{0^j1|\pst{X}}
   = \frac{F(k+j)}{\Phi(k)} \sim \frac{\alpha}{k}\left(\frac{k+j}{k}\right)^{-\alpha-1}
  ~.
\end{align*}
Meanwhile, so long as $k\geq \ell$, the truncated prediction has the form:
\begin{align*}
   \Prob{\mu}{0^j1|0^\ell}
   & = \sum_{n=1}^\infty \frac{\Phi(n+\ell)}{\sum_p \Phi(p+\ell)}
   \frac{F(n+\ell+j)}{\Phi(n+\ell)}\\
   & = \frac{\Phi(\ell+j)}{\sum_p \Phi(p+\ell)}
   \sim \frac{\alpha-1}{\ell}\left(\frac{\ell+j}{\ell}\right)^{-\alpha}
  ~.
\end{align*}
Now, choose $0<C<\alpha-1$ and define:
\begin{align*}
   B = \left(1-\frac{C+1}{\alpha}\right)^{-1}
  ~.
\end{align*}
Then it can be checked straightforwardly that whenever $k>B\ell$, we have:
\begin{align*}
   \Prob{\mu}{1|0^\ell}&-\Prob{\mu}{1|\pst{X}} \\
   & \sim \frac{1}{\ell}\left[\alpha\left(1-\frac{\ell}{k}\right)-1\right] \\
   & > \frac{C}{\ell}
  ~.
\end{align*}
The probability that $k>B\ell$ is given by $\Phi(B\ell) =
B^{-\alpha}\ell^{-\alpha}$. Setting
$K=B^{-\alpha}/\sqrt{\braket{1|1}_{\beta,\gamma}}$ proves the theorem.
\end{ProProp}

Therefore, while every sequence $\pst{X}$ converges to zero conditioning error
at finite length, this convergence is not uniform, to such a degree that the
proportion of pasts that retain conditioning error of $1/\ell$ has a fat tail in
$\ell$. This is a matter of practical importance that is not cleanly expressed
in the big-$O$ expression of the conditioning error from Thm.
\ref{thm:kernel-bayes-rule}.

Poisson and renewal processes are only the first and second rungs of the
hierarchy of hidden semi-Markov processes \cite{Marz17b}. Loosely speaking
these are state-dependent renewal processes. The preceding Hilbert space
framework extends to this larger process class---a subject recounted elsewhere.

\section{Concluding Remarks}
\label{sec:conclusion}

Taken altogether, the results fill-in important gaps in the foundations of
predictive states, while strengthening those foundations for further
development, extension, and application. Previously, predictive states were
only examined in the context of hidden Markov models, their generalizations,
and hidden semi-Markov models. We provided a definition applicable to any
stationary and ergodic process with discrete and real-valued observations.
Furthermore, we showed that predictive states for these processes are learnable
from empirical data, whether through a direct method of partitioning pasts or
through indirect methods, such as the reproducing kernel Hilbert space. 

One important extension is to continuous-time processes. By exploiting the full
generality of Jessen's and Enomoto's theorems we believe this extension is
quite feasible. As long as the set of possible pasts and futures constitutes a
separable space, they should be expressible in the form of a countable basis,
to which these theorems may then be applied. The challenge lies in constructing
an appropriate and useful basis. We leave this for future work.

We described key properties of the space in which predictive states live.
However, predictive states are not merely static objects. They predict the
probabilities of future observations. And, once those observations are made,
the predictive state may be updated to account for new information. Thus,
predictive states provide the stochastic rules for their own transformation
into future predictive states. This dynamical process has been explored in
great detail in the cases where the process is generated by a finite hidden
Markov model---this is found in former work on the \eM and the mixed states of
HMMs. Understanding the nature of this dynamic for more general processes,
including how it makes contact with other dynamical approaches---such as,
stochastic differential equations in the continuous-time setting---also remains
for future work.

\section*{Acknowledgments}
\label{sec:acknowledgments}

We thank Nicolas Brodu and Adam Rupe for helpful comments and revisions.  We
also thank Alex Jurgens, Greg Wimsatt, Fabio Anza, David Gier, Kyle Ray,
Mikhael Semaan, and Ariadna Venegas-Li for helpful discussions. JPC
acknowledges the kind hospitality of the Telluride Science Research Center,
Santa Fe Institute, Institute for Advanced Study at the University of
Amsterdam, and California Institute of Technology for their hospitality during
visits. This material is based upon work supported by, or in part by, Grant
Nos. FQXi-RFP-IPW-1902 and FQXi-RFP-1809 from the Foundational Questions
Institute and Fetzer Franklin Fund (a donor-advised fund of Silicon Valley
Community Foundation), grants W911NF-18-1-0028 and W911NF-21-1-0048 from the
U.S. Army Research Laboratory and the U.S. Army Research Office, and grant
DE-SC0017324 from the U.S. Department of Energy.


\end{document}